\documentclass[11pt]{article}

\usepackage{amsmath,amsthm,nicefrac}
\usepackage{amsfonts, amstext}

\usepackage[compact]{titlesec}
\usepackage{fullpage}
\usepackage[font={small,it}]{caption}

\usepackage{setspace}
\usepackage{comment}

\usepackage{enumitem}
\usepackage[breaklinks]{hyperref}
\hypersetup{colorlinks=true,%
            citebordercolor={.6 .6 .6},linkbordercolor={.6 .6 .6},%
citecolor=blue,urlcolor=black,linkcolor=blue,pagecolor=black}

\usepackage[nameinlink]{cleveref}
\Crefname{algocf}{Algorithm}{Algorithms}
\crefname{algocfline}{line}{lines}
\Crefname{invariant}{Invariant}{Invariants}
\Crefname{claim}{Claim}{Claims}
\Crefname{subclaim}{Subclaim}{Subclaims}

\usepackage{epsfig}
\usepackage{amsthm,amssymb}
\usepackage{amsthm,amssymb}

\usepackage{mathrsfs}
\usepackage{xspace}
\usepackage{soul}
\usepackage{latexsym}
\usepackage{bbm}

\usepackage{framed}

\usepackage[dvipsnames]{xcolor}
\definecolor{DarkGray}{rgb}{0.66, 0.66, 0.66}
\definecolor{DarkPowderBlue}{rgb}{0.0, 0.2, 0.6}
\definecolor{fluorescentyellow}{rgb}{0.8, 1.0, 0.0}

\usepackage[ruled,vlined,linesnumbered,algonl]{algorithm2e}
\SetEndCharOfAlgoLine{}
\SetKwComment{Comment}{\footnotesize$\triangleright$\ }{}

\SetCommentSty{mycommfont}

\usepackage{thmtools,thm-restate}

\allowdisplaybreaks

\newcommand{\alert}[1]{{\color{red}#1}}
\newcounter{note}[section]
\renewcommand{\thenote}{\thesection.\arabic{note}}
\newcommand{\agnote}[1]{\refstepcounter{note}$\ll${\bf Anupam~\thenote:}
  {\sf \color{blue} #1}$\gg$\marginpar{\tiny\bf AG~\thenote}}

\sethlcolor{fluorescentyellow}

\newcommand{\initOneLiners}{%
    \setlength{\itemsep}{0pt}
    \setlength{\parsep }{0pt}
    \setlength{\topsep }{0pt}
}
\newenvironment{OneLiners}[1][\ensuremath{\bullet}]
    {\begin{list}
        {#1}
        {\initOneLiners}}
    {\end{list}}
  
\newtheorem{theorem}{Theorem}[section]
\newtheorem{prop}[theorem]{Proposition}
\newtheorem{lem}[theorem]{Lemma}

\theoremstyle{definition}
\newtheorem{definition}[theorem]{Definition}

\newtheorem{assumption}[theorem]{Assumption}

\theoremstyle{remark}

\newcommand{\Ber}{\operatorname{Ber}}
\newcommand{\Binom}{\operatorname{Binom}}
\newcommand{\Var}{\operatorname{Var}}
\newcommand{\poly}{\operatorname{poly}}
\newcommand{\Vol}{\operatorname{Vol}}

\newcommand{\EE}{\mathbb{E}}

\newcommand{\choosealg}{\textsc{ChooseJobs}\xspace}
\newcommand{\algname}{\textsc{StochFree}\xspace}
\newcommand{\alg}{\textsf{alg}}
\newcommand{\opt}{\textsf{opt}}
\newcommand{\Opt}{\opt}
\newcommand{\Alg}{\alg}

\title{Minimizing Completion Times for Stochastic Jobs\\ via Batched Free Times}
\author{Anupam Gupta \\ Carnegie Mellon \\ anupamg@cs.cmu.edu 
	\and Benjamin Moseley \\ Carnegie Mellon\\ moseleyb@andrew.cmu.edu
	\and Rudy Zhou \\ Carnegie Mellon \\ rbz@andrew.cmu.edu}

\begin{document}

\maketitle

\begin{abstract}
  We study the classic problem of minimizing the expected total
  completion time of jobs on $m$ identical machines in the setting
  where the sizes of the jobs are stochastic. Specifically, the size
  of each job is a random variable whose distribution is known to the
  algorithm, but whose realization is revealed only after the job is
  scheduled. While minimizing the total completion time is easy in the
  deterministic setting, the stochastic problem has long been
  notorious: all known algorithms have approximation ratios that
  either depend on the variances, or depend linearly on the number of
  machines. 

  We give an $\widetilde{O}(\sqrt{m})$-approximation for stochastic
  jobs which have Bernoulli processing times. This is the first
  approximation for this problem that is both independent of the
  variance in the job sizes, and is sublinear in the number of
  machines $m$.
  Our algorithm is based on a novel reduction from minimizing the
  total completion time to a natural makespan-like objective, which we
  call the \emph{weighted free time}. We hope this free time objective
  will be useful in further improvements to this problem, as well as
  other stochastic scheduling problems. 
\end{abstract}

\thispagestyle{empty}

\newpage
\setcounter{page}{1}

\section{Introduction}\label{sec_intro}

Consider the problem of scheduling $n$ jobs on $m$ identical machines
to minimize the \emph{total completion time} of the jobs. If we assume
the job lengths are known, we can solve the problem optimally using
the \emph{shortest processing time} (SPT)
algorithm~\cite{DBLP:journals/cacm/BrunoCS74}. But what if the jobs durations are
not known precisely? In practical scheduling settings, the job sizes
are typically unknown. However, we can often give good stochastic
predictions based on jobs features and past data. In this work, we consider the setting where the job are
\emph{stochastic}, so the processing time of each job $j$ is an
independent random variable $X_j$ which is distributed according to a
\emph{known} probability distribution $\pi_j$, but whose
\emph{realized value we observe only after scheduling it} irrevocably
on some machine. Now the completion time $C_j$ of job $j$ is a random
variable, which depends on the random job sizes (and on any random
decisions our algorithm may make). Our goal is to minimize
$\sum_j \EE[C_j]$, the sum of expected completion times $C_j$ of the
jobs (or equivalently, their average).

Since the randomness in the job sizes is revealed as they are
scheduled, the decision of which job to schedule next (and on which
machine) can depend on the outcomes of already-scheduled jobs. Such
scheduling policies are called \emph{adaptive}. Formally, for each idle
machine, the \emph{adaptive scheduling policy} must choose which job
to schedule next on this machine---or it may choose to idle the
machine for some time period. In making this decision, it is allowed
to use any information it has gained from previously-scheduled
jobs. In particular, the policy knows the sizes $X_{j}$ of all
completed jobs $j$, and if a job $j$ has currently been run for $\tau$
time the policy knows that the jobs size is distributed as
$(X_j \mid X_j \geq \tau)$.

In this work we want to find near-optimal adaptive schedules, ones
that result in the total expected completion time being close to that achieved by the optimal adaptive schedule. Note this is an
apples-to-apples comparison where we relate the performance of our
solution to the best solution of the same type, and not to a
clairvoyant optimum that knows the future.
This problem has been of significant interest in both the theoretical
computer science and operations research communities for almost three
decades now~\cite{10.2307/3212936,
  10.2307/3214023,DBLP:journals/jacm/MohringSU99, DBLP:journals/siamcomp/SkutellaU05, DBLP:conf/stacs/JagerS18, DBLP:conf/cocoa/Schulz08,  DBLP:journals/mor/SkutellaSU16,
  DBLP:journals/mor/GuptaMUX20, DBLP:conf/stacs/ImMP15,
  DBLP:journals/orl/EberleFMM19}. 

While the deterministic problem can be solved optimally (using the
shortest processing time policy), the stochastic setting is
significantly more challenging. Early results for stochastic
completion time minimization focused on giving optimal policies only
for restricted classes of instances, e.g., the case where all job
distributions were exponentials, or where the jobs could be
stochastically ordered~\cite{10.2307/3212936, 10.2307/3214023}. Then,
starting with the ground-breaking work of M\"ohring, Schulz, and
Uetz~\cite{DBLP:journals/jacm/MohringSU99}, approximation algorithms
were given that worked for all stochastic instances. However, almost all such
algorithms have approximation ratios with at least linear dependence
on the \emph{squared coefficient of variation}
$\Delta := \max_j \frac{\Var(X_j)}{(\mathbb{E} X_j)^2}$
\cite{DBLP:journals/jacm/MohringSU99, DBLP:journals/siamcomp/SkutellaU05, DBLP:conf/stacs/JagerS18, DBLP:conf/cocoa/Schulz08,  DBLP:journals/mor/SkutellaSU16,
  DBLP:journals/mor/GuptaMUX20}. Since 
this squared coefficient of variation could be very large in general
(even for Bernoulli jobs), we want to obtain
approximations which are \emph{distribution-independent}, and in
particular, do not depend on the coefficient of variation.

There are significant roadblocks to obtaining such
distribution-independent guarantees: the known algorithmic toolkit for
deterministic jobs relies on greedy policies and linear program-based
algorithms \cite{DBLP:journals/mor/HallSSW97, DBLP:reference/crc/PruhsST04, 10.5555/1477600}. For the former, the natural
\emph{Shortest Expected Processing Time} (SEPT) policy has an approximation
ratio no better than $\Omega(n^{1/4})$
\cite{DBLP:conf/stacs/ImMP15}. Moreover, even for instances consisting
of only two types of jobs---identical unit-sized deterministic jobs
and identical Bernoulli jobs $X_j \sim s \cdot \Ber(p)$, no
\emph{index policy} (which assigns each job an ``index'' depending
only on its size distribution, and then schedules the jobs in order of
their indices) can have bounded approximation ratio
\cite{DBLP:journals/orl/EberleFMM19}. Approaches based on linear
programming also do not seem to extend to the stochastic setting: the
most expressive time-indexed linear program commonly used for such
settings has an integrality gap of $\Omega(\Delta)$
\cite{DBLP:journals/mor/SkutellaSU16}. Finally, we know that
\emph{adaptivity gap}---the gap between the optimal adaptive and fixed-assignment
policies\footnote{Such a policy non-adaptively assigns jobs to machines and runs each machines' jobs in a fixed order.}---is $\Omega(\Delta)$ \cite{DBLP:journals/mor/SkutellaSU16}.
Taken together, these lower bounds rule out most of the tools that
work for the setting of deterministic jobs.

The only distribution-independent approximations for stochastic
completion time minimization have approximation ratios at least linear
in $m$~\cite{DBLP:conf/stacs/ImMP15,
  DBLP:journals/orl/EberleFMM19}. In fact, nothing better than an
$O(m)$ approximation is known even for instances consisting of only
two types of jobs: identical unit-sized deterministic jobs and
identically distributed Bernoulli jobs $X_j \sim s \cdot \Ber(p)$
\cite{DBLP:journals/orl/EberleFMM19}. For general job distributions,
the best known approximation is $O(m \poly\log n)$
\cite{DBLP:conf/stacs/ImMP15}.


Again, there are barriers to obtaining approximations that are
sublinear in $m$: these previous works use ``volume'' lower bounds,
which rely on the fact that the processing capacity of $m$ machines is
$m$ times larger than that of a single machine. Indeed, the objective
is extremely sensitive to the number of machines $m$: decreasing the
number of machines from $m$ to $\frac{m}{2}$ can change the
solution's objective value by an exponential in $m$ factor. (This is
in stark contrast to the deterministic setting, where the optimal
solution's objectives for $m$ and $\frac{m}{2}$ machines differ by at
most a \emph{constant factor}.) See \Cref{sec_appendix_num_machines} for proofs of these two claims. This gives a sense for
why lower bounds on the optimal objective value based on the number of
machines $m$ do not generalize well to the stochastic setting, except
with a loss of a factor of $m$. 

In summary, the main question we ask is:
\begin{quote}
  \emph{Can we break through both the $\Delta$- and $m$-barriers for
    the basic problem of completion time minimization for stochastic jobs?}
\end{quote}
Despite the difficulty in obtaining improved approximations for this
problem, it is possible that the problem has a constant-factor
approximation!

\subsection{Our Results}

In this paper, we consider the case of (non-identical) \emph{Bernoulli
  jobs}, i.e., with independent processing times
$X_j \sim s_j \cdot \Ber(p_j)$ for size $s_j \geq 0$ and
probability $p_j \in [0,1]$. Our main result is the first algorithm
that is both distribution-independent and
has an approximation ratio sublinear in~$m$. We use the notation $\widetilde{O}(\cdot)$ to hide $\poly \log n$-factors.

\begin{theorem}[Main Theorem]
  \label{thm_mainber}
  There exists an efficient deterministic algorithm for
   completion time minimization for Bernoulli jobs that
  computes a list schedule that $\widetilde{O}(\sqrt{m})$-approximates the
  optimal adaptive policy.
\end{theorem}


By list schedule, we mean our algorithm produces a list (i.e., an ordering) of
all the jobs, and whenever a machine is free, it schedules the next
job according to this ordering.
Bernoulli jobs already are a significant generalization of the setting of \cite{DBLP:journals/orl/EberleFMM19},
and our result improves (up to $\poly\log n$-factors) the $O(m)$-approximation
of \cite{DBLP:journals/orl/EberleFMM19} and the
$\widetilde{O}(m)$-approximation of \cite{DBLP:conf/stacs/ImMP15} for this
special case of Bernoulli jobs. A corollary of our result is an
upper-bound of $\widetilde{O}(\sqrt{m})$ on the adaptivity gap between the
optimal adaptive policy and list schedules for the special case of
Bernoulli jobs.

We view Bernoulli jobs as an important testbed for new techniques for
this central problem in stochastic scheduling. Progress on this
problem has stalled since the the $\tilde{O}(m)$-approximation of
\cite{DBLP:conf/stacs/ImMP15}, and our work gives a new technical
framework (based on the new proxy objective function of free time
minimization) towards a $o(m)$-approximation for general
distributions.  We remark that a consequence of our techniques is a simple $\tilde{O}(m)$-approximation for Bernoulli jobs, matching (up to $\poly \log n$-factors) the result of \cite{DBLP:conf/stacs/ImMP15} in this special case (see \Cref{subsec_m} for details.)


Considering Bernoulli jobs has been an important
stepping-stone in other stochastic problems (e.g., for stochastic
makespan minimization \cite{DBLP:journals/siamcomp/KleinbergRT00}),
where algorithms for Bernoulli jobs could be extended to general
distributions. While we do not see how to get the extension yet, we
hope that our technical framework will soon lead to such extensions via our new proxy objective -- weighted free time (which is valid for general distributions) -- and the techniques we develop to optimize it.

\subsection{Technical Overview}


Our algorithm design will be informed by a proxy objective function, which we call the \emph{weighted free time}. We first observe that to bound the completion time of a job, it suffices to bound its starting time, $S_j$. This is because on identical machines, we have $C_j = S_j + X_j$, where $\sum_j X_j$ is a lower-bound on the optimal total completion time. The key idea of our proxy objective is to relate the per-job starting times to a more global quantity, which we call the \emph{free time.}

\begin{definition}[Free Time]
	Consider any fixed schedule. The $i$th free time of the schedule, which we denote by $F(i)$ is the first time when $i$ jobs have been started and at least one machine is free to start the $(i+1)$st job.
\end{definition}
 
 For schedules that do not idle machines, the $i$th free time is the load of the least-loaded machine after starting $i$ jobs. By definition of free time, there are $\Theta(n/2^k)$ jobs with starting times in $[F(n- n/2^{k-1}), F(n - n/2^k)]$ for all $k = 1, \dots, \log n$ (all logarithms are base $2$ in this paper.) Thus we have:
 \begin{gather}
   \sum_j S_j = \sum_{k = 1}^{\log n} \Theta(n/2^k) F(n - n/2^k). \label{eq:1}
 \end{gather}

We call this final expression, $\sum_{k = 1}^{\log n} n/2^k \cdot F(n - n/2^k)$, the \emph{weighted free time} of the schedule. We can view this objective as defining $\log n$ work checkpoints for our algorithm. These checkpoints are the time that we have $n/2^1$ jobs left to start (i.e. $F(n-n/2^1)$), $n/2^2$ left to start (i.e. $F(n-n/2^2)$), and so on. Roughly, the goal of our algorithm is to ensure that at each work checkpoint, our free time is comparable with the optimal schedule's free time at the same checkpoint.

We can now illustrate the reason for considering free times rather than the completion
time directly. Indeed, let $C(i)$ be the time that we complete $i$
jobs (and note the difference with $C_j$, which is the time at which we
finish a specific job $j$). We analogously have $\sum_j C_j =
\Theta(\sum_k n/2^k \cdot C(n-n/2^k))$. However, one difficulty of
stochastic jobs is we cannot easily control what are the first
$n-n/2^k$ jobs \emph{to complete}. On the other hand, for free times,
we have complete control over what $n-n/2^k$ jobs we decide \emph{to
  start} first, which then contribute to $F(n-n/2^k)$. This suggests two natural subproblems for our algorithm design:
\begin{itemize}
    \item \textbf{Subset Selection:} Compute nested sets of jobs $J_1 \subset J_2 \subset \dots$ such that for all $k$, $J_k$ is comparable to the first $n - n/2^k$ jobs of the optimal adaptive policy (i.e. the jobs contributing to $F(n-n/2^k)$ for $\opt$.)
    \item \textbf{Batch Free Time Minimization:} Given nested sets of jobs $J_1 \subset J_2 \subset \dots$, schedule the $J_k$'s such that our free time after scheduling $J_k$ is comparable to $F(n-n/2^k)$ for $\opt$. Our schedule must satisfy the \emph{batch constraint} that we schedule $J_k$ before $J_{k+1} \setminus J_k$ for all $k$.
\end{itemize}

The main technical challenge in both subproblems is the interaction
between the free time and the batch constraint. Since our final
algorithm will be a list schedule, the $J_k$-sets are chosen
non-adaptively. However, the optimal policy chooses its first
$n-n/2^k$ jobs \emph{adaptively}, so it is not clear that there even
exist good sets $J_k$. Our first contribution is that we can indeed
efficiently find good $J_k$ sets non-adaptively by delaying slightly
more jobs than $\opt$. Our algorithm will rely on a structural
characterization of the optimal adaptive policy for Bernoulli jobs. 

\begin{theorem}[Subset Selection, Informal]\label{thm_subset_selection_inf}
  Given Bernoulli jobs, we can efficiently find nested sets of jobs $J_1 \subset \dots \subset J_{\log n}$ such that $\lvert J_k \rvert = n - \widetilde{O}(n/2^k)$ and $J_k$ is a subset of the first $n - n/2^k$ jobs of the optimal adaptive policy for all realizations.
\end{theorem}

For a moment, suppose that we could schedule these jobs
$J_1 \subset \dots J_{\log n}$ \emph{optimally} (subject to the batch
constraint) to minimize the weighted free time. At first glance, it
might seem like we are done, because $J_k$ is always a subset of
$\opt$'s first $n-n/2^k$ jobs, so we are only scheduling fewer jobs
than $\opt$ at every work checkpoint. However, this reasoning fails
because of the batch constraint. Indeed, let us contrast the classic
makespan minimization problem (schedule \emph{deterministic} jobs to
minimize the load of the most-loaded machine) with its free time
analogue (schedule $n$ deterministic jobs to minimize the $n$th free
time). While an arbitrary list schedule is $O(1)$-approximate for
makespan, it is $\Omega(m)$-approximate for free time:

\begin{lem}\label{lem_freelower}
	For all $m > 1$, there exists a set of $n$ jobs $J$ and a list-schedule of $J$ whose $n$th free time $\Omega(m)$-approximates the optimal $n$th free time.
\end{lem}
\begin{proof}
  Consider $m$ ``small'' jobs of size $1$ and $m-1$ ``big'' jobs of size
  $m$. The optimal free time schedule is to first schedule one small
  job on each machine, and then one big job on $m-1$ machines. Thus
  the optimal free time is $1$.  Now consider the list-schedule of all
  big jobs before small jobs. Then each big job is scheduled on a
  separate machine, and all $m$ small jobs are scheduled on the
  remaining machine. This gives free time $m$.
\end{proof}

The instances from the above lemma suggest that we should schedule small jobs before bigger ones so that the big jobs do not clog up the machines and delay the starting times of the small jobs. 
In particular, it could be the case that $\opt$ does some small jobs
in $J_k \setminus J_{k-1}$ much earlier than our algorithm (due to the
batch constraint) when fewer machines are clogged by big
jobs. Further, the situation is more complicated because of stochastic
jobs. Consider an instance with two types of jobs: deterministic jobs
of size $1$ and Bernoulli jobs $X_j \sim s \cdot Ber(p)$ with $s \gg
1$ and $p \ll 1$. How should we determine which jobs are small or big?
More generally, there are intricate trade-offs between the sizes and
the probabilities of the jobs.

To overcome these technical challenges, roughly we show that our
$J_k$-sets are large enough (they do not delay too many small jobs) in
the following sense: Suppose we moved a small job from a later batch,
so $J \setminus J_k$, into $J_k$ because fewer machines are clogged
when scheduling $J_k$. However, by putting more jobs in an earlier
batch, we delay the free times of later batches. We show how to choose the $J_k$'s to maintain a delicate balance between the number of jobs delayed by $J_k$ and the number of clogged machines due to $J_k$. This ensures that there is not much benefit to moving small jobs to earlier batches. To achieve this, we initiate a systematic study of the free time.

In light of \Cref{thm_subset_selection_inf}, it remains to compute
good list schedule of the $J_k$'s subject to the batch constraint. We
will show that for our particular batches, it suffices to look at the
size parameter $s_j$ of our Bernoulli jobs to determine if a job is
small or big. 

\begin{theorem}[Batch Free Time Minimization, Informal]\label{thm_batch_ft_inf}
    Given the nested sets of Bernoulli jobs $J_1 \subset \dots \subset J_{\log n}$ guaranteed by \Cref{thm_subset_selection_inf}, list scheduling them in increasing order of size parameter (subject to the batch constraint) is $\widetilde{O}(\sqrt{m})$-approximate for weighted free time.
\end{theorem}

Combining \Cref{thm_subset_selection_inf,thm_batch_ft_inf} gives our
desired $\widetilde{O}(\sqrt{m})$-approximation for completion time
minimization for Bernoulli jobs.

\subsection{Comparison to prior work}

Prior $O(\Delta)$-approximations rely on bounding with respect to an LP solution, e.g. \cite{DBLP:journals/jacm/MohringSU99,DBLP:journals/mor/SkutellaSU16}. These have an integrality gap of $\Omega(\Delta)$. On the other hand, our algorithm is combinatorial and avoids this gap by comparing directly to the optimal adaptive policy.

The distribution-independent approximation of \cite{DBLP:conf/stacs/ImMP15} also partitions jobs into batches (as in our subset selection problem). Roughly, they guarantee that their batches are ``better'' than the optimal solution's jobs ``in expectation.'' However, we will show that our algorithm’s batches are better than the optimal solution's jobs for \emph{every} realization (using our structural characterization of the optimal policy \Cref{lem_sortprob}.)

Further, their algorithm schedules jobs within batches in arbitrary order (i.e. they give a trivial solution to the subproblem we call free time minimization). It seems likely that a loss of $\Omega(m)$ is necessary if one considers an arbitrary list schedule because of the lower bound in \Cref{lem_freelower}. To overcome this, we choose a particular list schedule (i.e. in increasing order of size parameter), which we show is $\widetilde{O}(\sqrt{m})$-approximate. To summarize, using a deeper technical understanding, we give more refined guarantees for both subset selection and free time minimization than \cite{DBLP:conf/stacs/ImMP15}.

The only other work is \cite{DBLP:journals/orl/EberleFMM19}, which considers even more restricted instances: those with only two types of jobs, identical deterministic and identical Bernoulli. Their algorithm is to schedule either all deterministic jobs first or all Bernoullis first (depending on the relative number of each type of jobs.) Our subset selection algorithm vastly generalizes this idea to arbitrary Bernoulli jobs with varying size and probability parameters. Further, while our algorithm in \Cref{thm_batch_ft_inf} runs an
index policy within each $J_k \setminus J_{k-1}$-batch, we
overcome the lower bound on index policies due to
\cite{DBLP:journals/orl/EberleFMM19} because our subset selection
algorithm constructs the batches by taking into account the relative
number of different types of jobs---not only the distributions of
individual jobs.

\subsection{Related Work}

Many stochastic combinatorial optimization problems have been studied from an approximation perspective; the previous results closest to this work are packing problems including those on stochastic versions of knapsack \cite{DBLP:conf/focs/DeanGV04, DBLP:conf/focs/GuptaKMR11, DBLP:conf/soda/BhalgatGK11, DBLP:conf/stoc/LiY13}, orienteering \cite{DBLP:conf/soda/GuptaKNR12}, multi-armed bandits \cite{DBLP:journals/jacm/GuhaMS10, DBLP:conf/soda/Ma14}, generalized assignment \cite{DBLP:conf/approx/AlaeiHL13}, and packing integer programs \cite{DBLP:conf/soda/DeanGV05}. Some stochastic versions of covering problems include $k$-TSP \cite{DBLP:conf/fsttcs/EneNS17, DBLP:conf/innovations/JiangLL020} and submodular cover \cite{DBLP:conf/soda/AgarwalAK19, pmlr-v139-ghuge21a}. Another important class of stochastic problems is probing/selection problems \cite{DBLP:journals/talg/GoelGM10, DBLP:conf/soda/GuptaNS16, DBLP:conf/soda/GuptaNS17, fu_et_al:LIPIcs:2018:9060}.

For stochastic scheduling problems, approximations are known for load balancing \cite{DBLP:conf/focs/GoelI99, DBLP:journals/siamcomp/KleinbergRT00, DBLP:conf/soda/Gupta0NS18, DBLP:conf/icalp/DeKLN20} and completion time minimization with precedence constraints \cite{DBLP:conf/soda/SkutellaU01}, preemption \cite{DBLP:journals/mor/MegowV14}, release dates and online arrivals \cite{DBLP:journals/mor/MegowUV06}; however the latter works have approximations that depend on the variance of job sizes.  

\section{Subset Selection}\label{sec_subset_selection}

The goal of this section is to solve the \emph{subset selection
  subproblem} for Bernoulli jobs: we want to find nested sets of jobs
$J_1 \subset \dots \subset J_{\log n}$ such that $J_k$ is comparable
to the first $n-n/2^k$ jobs of the optimal adaptive completion time
schedule. Formally, let $J_k^*$ be the random set consisting of the
first $n - n/2^k$ jobs scheduled by the optimal completion time
schedule. Our main theorem here is the following:

\begin{theorem}[Subset Selection]\label{lem_batch_prefix}
  Let $L$ be the number of distinct Bernoulli size parameters. There
  is an algorithm \choosealg that outputs sets $J_k$ satisfying:
  \begin{OneLiners}
  \item[(i)] $J_1 \subset \dots \subset J_{\log n} \subset J$
  \item[(ii)] $\lvert J_k \rvert \in [n - L \cdot n/2^k, n - n/2^k]$
  \item[(iii)] $J_k \subset J_k^*$ for all $k$ and all realizations.
  \end{OneLiners}
\end{theorem}
We show later how to use standard rescaling and discretization
techniques to assume $L = O(\log n)$ while losing only an extra
constant factor in our final approximation ratio. 

It is convenient to think of the $n/2^k$ jobs that the optimal
schedule \emph{excludes} from $J^*_k$ rather than the jobs it chooses
to start. Similarly, we specify our algorithm's set of jobs also by
the exclusions. The next lemma gives our structural characterization
for the optimal adaptive completion time schedule for Bernoulli jobs,
which allows us to characterize which jobs the optimal schedule
chooses to exclude.

Because jobs are Bernoullis, upon scheduling a job $j$, the scheduler immediately learns the realized size of $X_j$ because it is either $0$ or $s_j$. Thus, the optimal schedule can be represented by a decision tree, where each node is labeled by a job $j$, corresponding to the decision to schedule $j$ on the currently least loaded machine, and has a left- and right child, corresponding to the realized size of $j$ being $0$ or $s_j$, respectively. Every root-leaf path on this tree gives an ordering to schedule the jobs for a particular realization of job sizes.

\begin{restatable}{lem}{exchange}\label{lem_sortprob}
    Consider a collection of Bernoulli jobs. Then for each possible size parameter, the optimal adaptive completion time schedule for these jobs starts the jobs with this size parameter in increasing order of their probabilities for all realizations of the job sizes.
\end{restatable}
\begin{proof}[Proof Sketch]
    Our proof is an exchange argument. Consider the optimal decision tree (as described above), and suppose there exists a root-leaf path that schedules job $X_b \sim s \cdot \Ber(p_b)$ before $X_a \sim s \cdot \Ber(p_a)$ with $p_a \leq p_b$. Then there exists a subtree $T$ rooted at $b$ such that $a$ is scheduled on every root-leaf path in this subtree.
    
    We now show how to modify $T$ to start $a$ before $b$ while not
    increasing the expected completion time. Let $T_L$ and $T_R$ be
    the left- and right subtrees (corresponding to the root job $b$
    coming up size $0$ or $s$) of $T$, respectively. We define $T_L(a
    \rightarrow b)$ to be $T_L$ with the job $a$ replaced by job $b$
    and $T_L(-a)$ to be $T_L$, but at $a$'s node, we do not schedule
    anything and instead go to $a$'s left child. The subtrees $T_R(a
    \rightarrow b)$ and $T_R(-a)$ are defined analogously. See
    Figure~\ref{fig:exch} for the modified tree $T'$.
    
        \begin{figure}
        \centering
        \includegraphics[width=0.75\textwidth]{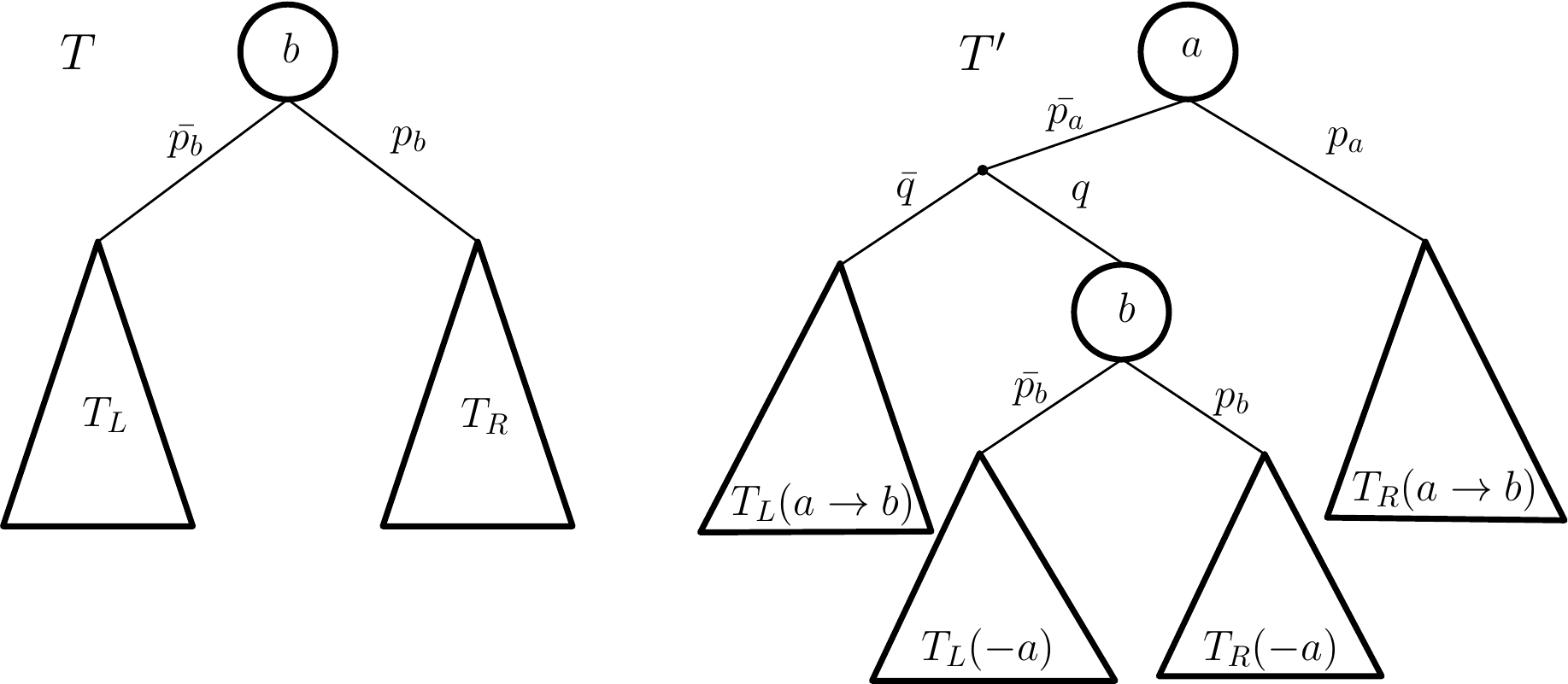}
        \caption{Original decision tree $T$ and modified decision tree
        $T'$}
        \label{fig:exch}
    \end{figure}

    We choose the parameter $q$ so that the probability that $T'$
    enters $T_L(a \rightarrow b)$ or $T_L(-a)$ is exactly
    $\bar{p}_b$. This is our replacement for the event that $T$ enters
    $T_L$. A calculation now shows that the expected completion time
    weakly decreases from $T$ to $T'$. (See the proof in
    \Cref{sec_appendix_exchange} for details.) 
\end{proof}

By definition $\opt$ can exclude only $n/2^k$ jobs from $J_k^*$. On the other hand, our algorithm will exclude $n/2^k$ jobs of
each Bernoulli size \textit{simultaneously}. \Cref{lem_sortprob} suggests that we might
as well exclude the jobs with largest $p_j$'s. In particular, our
algorithm to choose sets of jobs that are comparable to the $J_k^*$'s
is the following:

\medskip
\hrule
\smallskip
\hrule
\smallskip
\choosealg: For each $k = 1, \dots, \log n$, let $J_k$ be the set of jobs constructed as follows:
    	\begin{OneLiners}
    		\item[i.] Initialize $J_k = J$.
    		\item[ii.] For each Bernoulli size $s$, remove from $J_k$ the $n/2^k$ jobs of size $s$ with largest $p_j$'s.
    		\item[iii.] Output $J_k$.
        \end{OneLiners}
\medskip
\hrule
\smallskip
\hrule
\medskip


\begin{proof}[Proof of~\Cref{lem_batch_prefix}]
    	It is immediate that the sets $J_1, \dots, J_{\log n}$ are nested and have the desired size. It remains to show that $J_k \subset J_k^*$. Note that $\opt$ excludes at most $n/2^k$ jobs of the largest probabilities of each size by \Cref{lem_sortprob}. This holds for all realizations. On the other hand, we exclude $n/2^k$ jobs of largest probability of all sizes simultaneously.
\end{proof}

Morally, \Cref{lem_batch_prefix} states that we know the jobs that $\opt$
starts to achieve $F^*(n-n/2^k)$ for all $k$ (up to a $L$-factor.)
This suggests that we should first schedule $J_1$ to get free time
comparable to $F^*(n-n/2^1)$, and then $J_2 \setminus J_1$ to get free
time comparable to $F^*(n-n/2^2)$, and so on.

The goal of the next section is to show how to schedule the $J_k$'s
subject to this batch constraint (we must schedule all jobs in
$J_{k-1}$ before any in $J_k \setminus J_{k-1}$ for all $k$) such that
our weighted free time is comparable to that of $\opt$. Note that in
general, even though $J_k \subset J_k^*$ for all $k$, the
\emph{optimal} completion time schedule may not satisfy the batch
constraint---this is precisely is the main technical challenge that
we have to overcome in the next section.


\section{Batch Free Time Minimization}\label{sec_batch_free_time}

We now turn to the \emph{batch free time minimization} problem. Our starting point is the nested sets of jobs $J_1 \subset \dots \subset J_{\log n} \subset J$ output by \choosealg. Recall that $J_k^*$ is the random set of the first $n-n/2^k$ jobs scheduled by the optimal completion time policy $\Opt$.

\subsection{Free Time Basics}
\label{sec:free-time-basics}

To motivate our final algorithm, we explore some basic properties of the free time. We first recall the lower bound instance from \Cref{lem_freelower}: there are $m$ small jobs of size $1$ and $m-1$ big jobs of size $m$. The optimal free time is $1$ by first scheduling all small jobs and then all big jobs. Observe that the final $m-1$ big jobs do not contribute to the optimal free time. This gave the intuition that we should schedule small jobs before big ones clog up the machines. This intuition turns out to be correct, which we formalize in the next lemma.

\begin{lem}\label{lem_singlefree}
	List-scheduling deterministic jobs in increasing order of size is a $4$-approximation for free time.
\end{lem}
\begin{proof}
  Let $J$ be the set of input jobs and $\Opt$ the optimal free
  time. We partition $J$ into \emph{small} and \emph{big} jobs: a job
  is big if its size is strictly greater than $\Opt$, and is small
  otherwise. This definition means that $\Opt$ can schedule at most one big job per machine. Moreover, there are strictly less than $m$ big jobs, otherwise $\Opt$ would need to schedule at least one big job per machine, which contradicts the optimal free time.
	
	On the other hand, our algorithm starts all small jobs before all big jobs. We claim that $F(J)$ is at most the makespan of our algorithm after only scheduling the small jobs, which we denote by $M(S)$. To see this, consider the time right before we start the first big job. All machines have loads within $[M(S) - \Opt, M(S)]$ (the lower bound follows by noting that we list-scheduled only jobs of size at most $\Opt$ up until this point.) Thus, we schedule the at most $m-1$ remaining big jobs each on separate machines. All of the big jobs are started by time $M(S)$, and there exists a machine that schedules no big job, which is free by $M(S)$ as well. It follows, $F(J) \leq M(S)$.
	
	Now, because every list-schedule is $2$-approximate for makespan, we have $M(S) \leq 2 M^{\Opt} (S)$, where $M^{\Opt}(S)$ is the makespan of the small jobs under $\Opt$ (i.e. when $\Opt$ finishes its final small job.) To complete the proof, we relate $M^{\Opt}(S)$ with $\Opt$. It suffices to show $M^{\Opt}(S) \leq 2 \Opt$. To see this, consider time $\Opt$ in the optimal schedule. At this time, all machines are either free or working on their final job. In particular, any machine working on a small job completes by time $\Opt + \Opt$.
\end{proof}

While we do not apply \Cref{lem_singlefree} directly for our algorithm, the ideas in the analysis will be crucial. In particular, a key concept in our analysis is to differentiate between small and big jobs. Roughly, when we consider $J_k$, a job is small if its size is at most $F^*(n-n/2^k)$ and big otherwise. We are concerned about the volume (total size) of the small jobs and number of big jobs. However, because of the batch constraint, we cannot ensure that all small jobs are scheduled before all big jobs in general.

As we schedule batches $J_1, J_2 \setminus J_1, \dots, J_k \setminus J_{k-1}$, more and more machines are getting clogged by big jobs. For the purposes of scheduling $J_k$, these machines are effectively turned off. Thus, as we proceed through the batches, we are averaging the volume of small jobs over fewer and fewer machines. The goal of our algorithm will be to ensure that we do small jobs as early as possible (subject to the batch constraint) so that we have the most unclogged machines available. 

\subsection{Final Algorithm}
\label{sec:final-algorithm}

With the goal of \S\ref{sec:free-time-basics} in mind, we are ready to
describe our final algorithm, which is the
$\widetilde{O}(\sqrt{m})$-approximation guaranteed by
\Cref{thm_mainber}. Although we cannot ensure that within a batch,
jobs are scheduled in increasing order of \emph{realized} size as in
the analysis of \Cref{lem_singlefree}, because our jobs are
Bernoullis, we can ensure that all jobs that come up heads (have
realized size $s_j$) are scheduled in increasing order of realized
size. Here, we crucially use the fact that our jobs are Bernoullis, so
if the come up tails (have size $0$), they do not affect the free
time. For the rest of the description, we make the following assumption, which we justify in \Cref{sec_appendix_rescale}.

\begin{assumption}\label{as_rescale}
    We assume that there are $L = O(\log n)$ distinct Bernoulli size parameters $s_j$, each at most $n^8$.
\end{assumption}

By losing a constant factor in the final approximation ratio, we may
assume \Cref{as_rescale} for the rest of the analysis.

\begin{restatable}{lem}{rescale}\label{lem_bernoilli_rescale}
    Let $m \geq 2$. Suppose there exists an algorithm for completion time minimization for Bernoulli jobs on $m$ machines satisfying \Cref{as_rescale} that outputs a list schedule with expected completion time at most $\alpha \big(\mathbb{E} \opt + O(1)\big)$. Then there exists a $O(\alpha)$-approximate algorithm for the same problem without the assumption. Further, the resulting algorithm is also a list schedule, and it preserves efficiency and determinism.
\end{restatable}

The proof uses
standard rescaling and discretization ideas, but it is more involved
because of the stochastic jobs; we justify it in
\Cref{sec_appendix_rescale}. Our final algorithm is now the
following:

\medskip
\hrule
\smallskip
\hrule
\medskip
\algname: Given input collection $J$ of Bernoulli jobs:
\begin{itemize}
    \item[i.] Run \choosealg to obtain nested sets of jobs $J_1 \subset \dots \subset J_{\log n} \subset J$
    \item[ii.] List-schedule each batch $J_k \setminus J_{k-1}$ in increasing order of Bernoulli size parameter $s_j$ for all batches $k = 1, \dots, \log n$.
    \item[iii.] List-schedule all remaining jobs $J \setminus J_{\log n}$ in arbitrary order.
\end{itemize}
\hrule
\smallskip
\hrule
\medskip

It is clear that \algname outputs a list schedule in polynomial
time and is deterministic. Our main approximation guarantee for \algname is the following.

\begin{restatable}[Batch Free Time Minimization]{theorem}{mainalg}\label{thm_main_alg}
      Given Bernoulli jobs, if $m \geq 2$ and \Cref{as_rescale} holds,
  then \algname outputs a list schedule with expected completion time
  at most $\widetilde{O}(\sqrt{m}) \cdot \big(\mathbb{E}[\opt] + O(1)\big)$, where $\opt$
  is the optimal adaptive policy.
\end{restatable}

Note that composing \Cref{thm_main_alg} with \Cref{lem_bernoilli_rescale} gives the desired $\widetilde{O}(\sqrt{m})$ without the assumption for all $m \geq 2$. For the remaining case of $m=1$, scheduling the jobs in increasing
order of their expected processing times is an optimal policy \cite{doi:10.1287/mnsc.12.9.707}. This gives the desired $\widetilde{O}(\sqrt{m})$-approximation for all $m$, and completes the proof of \Cref{thm_mainber}. In the remainder of the paper, we analyze \algname (\Cref{thm_main_alg}.)



\section{Analysis of the \algname Algorithm}\label{sec_alg_analysis}

The goal of this section is to prove \Cref{thm_main_alg}, given
\Cref{as_rescale}. Our proof has four conceptual steps.

\begin{itemize}
    \item[i.] Bound the weighted free time of $\alg$ by averaging the
      volume of small jobs within each batch over the unclogged
      machines---those that have not yet scheduled a big job. (\S\ref{subsec_free_time}) 
    \item[ii.] Show that \algname is $\widetilde{O}(m)$-approximate
      for all $m \geq 2$. This serves as a warm-up to the improved
      $\widetilde{O}(\sqrt{m})$-approximation, and it allows us to
      focus on the remaining case where $m = \Omega(1)$ is
      sufficiently large. (\S\ref{subsec_m})
    \item[iii.] Control the rate that machines become clogged by a large job. We show that the rate that machines become clogged for $\alg$ is slow enough so that $\opt$ cannot benefit much by putting small volume in earlier batches than $\alg$. (\S\ref{subsec_bound_clog})
    \item[iv.] Finally, bound the contribution of the volume of small
      jobs to the free time. Here we handle the main challenge, which
      is that $\opt$ may schedule small volume ``in the past''
      compared to
      $\alg$. (\S\ref{subsec_small_past}-\ref{subsec_coin})
\end{itemize}

\subsection{Weighted free time}\label{subsec_free_time}

First, we pass from total completion time to our proxy objective of weighted free time. We let $\Opt$ denote the optimal adaptive completion time policy as well as its completion time. Recall that $J_k^*$ is the first $n-n/2^k$ jobs scheduled by this policy achieving free time $F^*(n-n/2^k)$. Analogously, we let $\Alg$ denote the completion time of our algorithm, and $F(J_k)$ the free time of our algorithm after scheduling $J_k$.

\begin{lem}\label{lem_new_obj}
    We have $\Alg = O(\log n) (\sum_k n/2^k \cdot F(J_k) + \Opt)$ and $\Opt = \Omega(\sum_k n/2^k \cdot F^*(n-n/2^k))$.
\end{lem}
\begin{proof}
    We rewrite $\Alg = \sum_j C_j = \sum_j S_j + \sum_j X_j$. First,
    note that $\sum_j X_j \leq \Opt$. It remains to bound $\sum_j
    S_j$. Recall that in \algname, first we list-schedule $J_{\log n}$ subject to the batch constraint and then $J \setminus J_{\log n}$.   
    We first handle the starting times of $J_{\log n}$. For all $k$,
    note that $J_k \setminus J_{k-1}$ consists of at most $O(L) \cdot
    n/2^k$ jobs with starting times in $[F(J_{k-1}), F(J_k)]$ by
    \Cref{lem_batch_prefix}. Thus we have $\sum_{j \in J_{\log n}} S_j
    = O(L)\cdot (\sum_k n/2^k \cdot F(J_k)) = O(\log n)\cdot (\sum_k n/2^k \cdot F(J_k))$ using \Cref{as_rescale}.
    
    For the jobs in $J \setminus J_{\log n}$, by \Cref{lem_batch_prefix}, there are at most $O(L) = O(\log n)$ such jobs. Each of these jobs completes by the makespan of $\Alg$'s schedule. Further, $\Alg$ is a list schedule, which is $2$-approximate for makespan, so the makespan of $\Alg$ is at most twice the makespan of $\Opt$. The makespan of $\Opt$ is a lower bound on $\Opt$ (because some job must complete at this time.) We conclude, $\sum_{j \in J \setminus J_{\log n}} S_j = O(\log n) \Opt$. Combining our bounds for $J_{\log n}$ and $J \setminus J_{\log n}$ gives the desired result for $\Alg$.
    
    The bound on $\opt$ follows from \Cref{eq:1}.
\end{proof}

We refer to $\sum_k n/2^k \cdot F(J_k)$ as $\Alg$'s weighted free time and $\sum_k n/2^k \cdot F^*(n-n/2^k)$ as $\Opt$'s. The remainder of the analysis will focus on bounding $\Alg$'s weighted free time with respect to $\Opt$'s. Our main result is the following:

\begin{restatable}{theorem}{freemain}\label{thm_free_main}
    If $m = \Omega(1)$ is sufficiently large, then the weighted free time of $\Alg$ satisfies:
    \[\mathbb{E}\bigg[ \sum_k n/2^k \cdot F(J_k)\bigg] =
      \widetilde{O}(\sqrt{m}) \cdot \bigg( \mathbb{E} \bigg[\sum_k n/2^k \cdot
      F^*(n-n/2^k)\bigg] + \mathbb{E}[\Opt\,] \bigg) + O(1).\]
\end{restatable}

Note that \Cref{thm_free_main} along with \Cref{lem_new_obj} implies the desired guarantee in \Cref{thm_main_alg} for the case $m = \Omega(1)$ sufficiently large.

We now introduce some notations. For all $k$, we call $I_k = J_k \setminus J_{k-1}$ the $k$th \emph{batch} of jobs. Recall that the $J_k$'s are nested, so the batch constraint says we schedule in order $I_1, \dots, I_{\log n}$. We define $I_k^* = J_k^* \setminus J_{k-1}^*$ analogously. For any set of jobs, $J'$ and $\tau \geq 0$, we define $J'(=\tau)$ to be the random subset consisting of all jobs in $J'$ with \emph{realized} size exactly $\tau$. We define $J'(>\tau)$ and $J'(\leq \tau)$ analogously. Further, for a set of jobs $J'$, we let $\Vol(J') = \sum_{j \in J'} X_j$ be the volume of $J'$. Finally, we say job $j$ is $\tau$-big for $\tau \geq 0$ if $X_j > \tau$. Otherwise $j$ is $\tau$-small.

As in the analysis for minimizing the free time for a single batch of deterministic jobs (\Cref{lem_singlefree}), the key concept is to differentiate between small and big jobs. To this end, for all $k$ we define the random threshold $\tau_k = 2 \cdot \max(\mathbb{E} F^*(n - n/2^k), F^*(n - n/2^k))$. Morally, one should imagine that $\tau_k$ is $F^*(n - n/2^k)$, but there is an edge case where $F^*(n - n/2^k) < \mathbb{E} F^*(n - n/2^k)$ and a multiplicative factor for concentration. When bounding $F(J_k)$, we will take $\tau_k$ to be our threshold between small- and big jobs. This threshold has the following crucial property that $\alg$ always has at least as many unclogged machines as $\opt$. In particular, $\alg$ always has at least one unclogged machine.
    
\begin{prop}\label{prop_big_threshold}
    For all $k$, the following holds per-realization: $\lvert J_k(> \tau_k) \rvert \leq \lvert J_k^*(> \tau_k) \rvert < m$.
\end{prop}
\begin{proof}
    The first inequality follows from \Cref{lem_batch_prefix}, because $J_k(> \tau_k) \subset J_k^*(> \tau_k)$ per-realization. For the second inequality, note that $\tau_k \geq F^*(n - n/2^k)$, so by definition of free time, $\Opt$ schedules strictly less than $m$ jobs bigger than $\tau_k$ to achieve $F^*(n-n/2^k)$. 
\end{proof}

Using this threshold, we re-write $F(J_k)$ by averaging the volume of small jobs over the unclogged machines (the ones with no big job.)

\begin{lem}\label{prop_k_free_vol}
    For all $k$, the following holds per-realization:
    \[F(J_k) \leq F(J_{k-1}) + \frac{\Vol(I_k(\leq \tau_k))}{m- \lvert J_{k-1}(> \tau_k) \rvert} + 2 \tau_k.\]
\end{lem}
\begin{proof}
    First, we note by \Cref{prop_big_threshold} that the denominator $m - \lvert J_{k-1}(> \tau_k) \rvert \geq 1$. Then consider time $F(J_{k-1})$. There are at least $m - \lvert J_{k-1}(> \tau_k) \rvert$ machines that have not scheduled a $\tau_k$-big job in $J_{k-1}$. At this time, each machine is either free or working on its final job in $J_{k-1}$. In particular, each machine that has not scheduled a $\tau_k$-big job in $J_{k-1}$ is free to start working on $I_k$ by time $F(J_{k-1}) + \tau_k$, and there are at least $m - \lvert J_{k-1}(> \tau_k) \rvert$ such machines.
    
    We need the following monotonicity property of list schedules.
    
    \begin{lem}\label{lem_listprop}
		Consider a set of deterministic jobs and a fixed list schedule of those jobs. Then increasing the initial load or decreasing the number of machines weakly increase the free time of the schedule.
	\end{lem}
	\begin{proof}
		Let $J$ be the set of jobs. Consider initial load vectors $\ell, \ell' \in \mathbb{R}^m$, where the $i$th entry of each vector denotes the initial load on machine $i$. Now suppose $\ell \leq \ell'$, entry-wise. It suffices to show that $F(J,\ell) \leq F(J,\ell')$, where $F(J,\ell)$ is the free time achieved by our list-schedule with initial load $\ell$. This suffices, because we can decrease the number of machines by making the initial loads of some machines arbitrarily large so that they will never be used.
		
		We prove $F(J,\ell) \leq F(J,\ell')$ by induction on the number of jobs, $\lvert J \rvert$.
		In the base case, $\lvert J \rvert = 0$, so the claim is trivial because $\ell \leq \ell'$.
		For $\lvert J \rvert > 0$, let $j$ be the first job in the list, which is scheduled, without loss of generality, on the first machine for both initial loads $\ell$ and $\ell'$. Then:
		\[F(J,\ell) = F(J \setminus \{j\}, \ell + s_j e_1) \leq F(J \setminus \{j\}, \ell' + s_j e_1) = F(J,\ell'),\]
		where $e_1$ is the first standard basis vector, so we have $\ell + s_j e_1 \leq \ell' + s_j e_i1$ entry-wise. Then we assumed inductively that $F(J \setminus \{j\}, \ell + s_j e_1) \leq F(J \setminus \{j\}, \ell' + s_j e_1)$.
	\end{proof}
    
    By \Cref{lem_listprop}, we can upper-bound $F(J_k)$ by list-scheduling $I_k$ with initial load $F(J_{k-1}) + \tau_k$ on $m - \lvert J_{k-1}(> \tau_k) \rvert$ machines that have not scheduled a $\tau_k$-big job in $J_{k-1}$. Recall that $\Alg$ list-schedules $I_k$ in increasing order of size parameter, so - ignoring jobs that come up tails with realized size $0$ - we schedule all $\tau_k$-small jobs in $I_k$ before any $\tau_k$-big one. Further, $\lvert I_k(>\tau_k)\rvert < m - \lvert J_{k-1}(> \tau_k) \rvert$ by \Cref{prop_big_threshold}, so there exists some machine that schedules only $\tau_k$-small jobs in $I_k$.  This machine is free by time $F(J_k) \leq F(J_{k-1}) + \tau_k + \frac{\Vol(I_k(\leq \tau_k))}{m- \lvert J_{k-1}(> \tau_k) \rvert} + \tau_k$.
\end{proof}

Using \Cref{prop_k_free_vol} and the exponentially decreasing weights, we can re-write $\Alg$'s weighted free time as:
\begin{equation}\label{eq_free_time_vol}
    \sum_k n/2^k \cdot F(J_k) = O\bigg(\sum_k n/2^k \cdot \frac{\Vol(I_k(\leq \tau_k))}{m- \lvert J_{k-1}(> \tau_k) \rvert} + \sum_k n/2^k \cdot \tau_k\bigg)
\end{equation}
By definition of $\tau_k$, the second sum is $O(\mathbb{E} \sum_k n/2^k \cdot F^*(n-n/2^k))$ in expectation, which is exactly $\Opt$'s weighted free time. It remains to bound the first sum. 

\subsection{Warm up: $\widetilde{O}(m)$-approximation}\label{subsec_m}


Before proceeding with the proof of \Cref{thm_free_main}, we observe
that \Cref{eq_free_time_vol} along with our basic weighted free time
properties is enough to give a
$\widetilde{O}(m)$-approximation. Interestingly, this gives a simple
proof that nearly matches the previously best-known guarantees for Bernoulli jobs.

\begin{lem}\label{lem_warm_up_m}
    Given Bernoulli jobs, if $m \geq 2$ and \Cref{as_rescale} holds,
  then \algname outputs a list schedule whose expected completion time $\widetilde{O}(m)$-approximates the optimal adaptive policy. 
\end{lem}
\begin{proof}
    Starting from \Cref{eq_free_time_vol}:
    \[\sum_k n/2^k \cdot F(J_k) = O\bigg(\sum_k n/2^k \cdot \frac{\Vol(I_k(\leq \tau_k))}{m- \lvert J_{k-1}(> \tau_k) \rvert} + \sum_k n/2^k \cdot \tau_k\bigg),\]
    we note that $I_k \subset J_k^*$ by \Cref{lem_batch_prefix} and $m - \lvert J_{k-1}(> \tau_k) \rvert \geq 1$ by \Cref{prop_big_threshold}. Thus, we can bound:
    \[\frac{\Vol(I_k(\leq \tau_k))}{m- \lvert J_{k-1}(> \tau_k) \rvert} \leq \Vol(J_k^*(\leq \tau_k)).\]
    We claim that $\Vol(J_k^*(\leq \tau_k)) = O(m) \cdot \tau_k$. To see this, observe that by averaging the volume of $J_k^*(\leq \tau_k)$ over the $m$ machines, after scheduling $J_k^*$, each machine in $\opt$ has load at least $\frac{\Vol(J_k^*(\leq \tau_k))}{m} - \tau_k$. This gives $F^*(n-n/2^k) \geq \frac{\Vol(J_k^*(\leq \tau_k))}{m} - \tau_k$. Noting that $\tau_k \geq F^*(n-n/2^k)$ completes the proof that $\Vol(J_k^*(\leq \tau_k)) = O(m) \cdot \tau_k$.
    
    Applying this to our above expression gives $\frac{\Vol(I_k(\leq \tau_k))}{m- \lvert J_{k-1}(> \tau_k) \rvert} = O(m) \cdot \tau_k$, so $\alg$'s weighted free time satisfies:
    \[\sum_k n/2^k \cdot F(J_k) = O(m) \cdot (\sum_k n/2^k \cdot \tau_k).\]
    Taking expectations and applying \Cref{lem_new_obj} completes the proof.
\end{proof}

The loss of $m$ in the above proof was because $\alg$ averages the
small volume over at least $1$ unclogged machine, but $\opt$ may
average the same volume over at most $m$ machines. Intuitively, this
reasoning is why previous work loses a $m$-factor as well. 

Further, this is the main technical challenge that we will overcome to
get our improvement. Indeed, even though $J_k \subset J_k^*$ for all $k$, it is not true that $I_k \subset I_{k^*}$. This means that while we are averaging $\Vol(I_k(\leq \tau_k))$ over $m - \lvert J_{k-1}(> \tau_k) \rvert$ machines (which is at least as many machines as $\Opt$ has for batch $k$), it can be the case that $\Opt$ actually did jobs in $I_k$ in much earlier batches. 
In the remainder of our analysis, we do a more fine-grained analysis of the rate that $\alg$ and $\opt$ clog machines, and when they choose to do the same small volume. This allows us to break through the linear dependence in $m$.

\subsection{Bounding the unclogged machines}\label{subsec_bound_clog}

In this section, we are interested in controlling the quantity $m - \lvert J_{k-1}(> \tau_k) \rvert$, which is the number of machines we have left to schedule $I_k$ (the unclogged machines.) Note that there are two sources of randomness: the realizations of jobs in $J_{k-1}$ and the threshold $\tau_k$.

Our strategy is to control $m - \mathbb{E} \lvert J_{k-1}(> \tau) \rvert$ for a fixed threshold $\tau$. Then because $\lvert J_{k-1}(> \tau) \rvert$ is a sum of independent $\{0,1\}$-valued random variables, a Chernoff-union bound argument allows us to control $m - \lvert J_{k-1}(> \tau) \rvert$ as well.

However, we will see that concentration alone is not enough; this is because there is an unbounded difference between $\lvert J_{k-1}(> \tau) \rvert = m$ and $\lvert J_{k-1}(> \tau) \rvert < m - 1$. In the former case, all machines are clogged by big jobs, whose size we cannot upper bound. Thus, we cannot make any progress towards reaching time $F(J_k)$ (by starting more jobs.) In the latter, we have at least one machine, so we can still make some progress towards $F(J_k)$. The situation to keep in mind is when $\mathbb{E} \lvert J_{k-1}(> \tau) \rvert$ is close to $m$, so concentration around the mean will fail to preserve this hard constraint that we need at least one unclogged machine. To remedy this, we will combine concentration arguments with the per-realization properties of \algname. 

We begin with the concentration arguments, so we wish to understand $m - \mathbb{E} \lvert J_{k-1}(> \tau) \rvert$. We first use the properties of \choosealg to bound $\mathbb{E} \lvert J_{k-1}(> \tau) \rvert$:

\begin{prop}\label{prop_batch_decrease}
    For all fixed thresholds $\tau$ and batches $k$, we have $\mathbb{E} \lvert I_k(>\tau) \rvert \geq \frac{1}{2} \mathbb{E\tau} \lvert I_{k-1}(>\tau)\rvert$.
\end{prop}
\begin{proof}
    By summing over the relevant sizes, it suffices to prove $\mathbb{E} \lvert I_k(=s) \rvert \geq \frac{1}{2} \mathbb{E} \lvert I_{k-1}(=s)\rvert$ for any Bernoulli size parameter $s$. We may assume $\mathbb{E} \lvert I_{k-1}(=s)\rvert > 0$ or else the proposition is trivial. 
    
    Then when \choosealg constructs $J_{k-1}$, it includes at least one job with size parameter $s$. It follows, there exist $n/2^{k-1}$ remaining jobs in $J \setminus J_{k-1}$ with size parameter $s$. When constructing $J_k$, \choosealg will include $n/2^k$ of these remaining jobs. In conclusion, $I_{k-1}$ has at most $n/2^{k-1}$ jobs with size parameter $s$, while $I_k$ has at least $n/2^k$. The result follows because \choosealg includes jobs in increasing order of $p_j$.
\end{proof}

\Cref{prop_batch_decrease} allows us to relate the expected number of machines left (with respect to fixed threshold $\tau$) at batch $k$ with the number of machines left at $k' \leq k$:

\begin{lem}\label{lem_machine_remain}
    For all fixed thresholds $\tau$ and batches $k' \leq k$, we have $m' - \mathbb{E}\lvert J_{k-1}(> \tau) \rvert \geq 2^{-(k-k' + 1)} \cdot (m' - \mathbb{E}\lvert J_{k'-1}(> \tau)\rvert)$, where $m' \geq \mathbb{E} \lvert J_k(>\tau) \rvert$.
\end{lem}
\begin{proof}
    We may assume $\mathbb{E} \lvert I_k(> \tau) \rvert > 0$ or else the lemma is trivial, because by definition of \choosealg, if $\mathbb{E} \lvert I_k(> \tau) \rvert = 0$, then $\mathbb{E} \lvert J_{k-1}(> \tau)\rvert = 0$ and $\mathbb{E} \lvert J_{k'-1}(> \tau)\rvert = 0$.
    
    In particular, we may assume $m' - \mathbb{E} \lvert J_{k-1}(> \tau) \rvert \geq \mathbb{E} \lvert I_{k}(> \tau) \rvert > 0$. Then we compute:
    \[\frac{m' - \mathbb{E} \lvert J_{k' -1}(> \tau) \rvert}{m' - \mathbb{E} \lvert J_{k -1}(> \tau) \rvert} = 1 + \frac{\mathbb{E} \lvert I_{k'}(> \tau) \rvert + \dots + \mathbb{E} \lvert I_{k-1}(> \tau) \rvert}{m' - \mathbb{E} \lvert J_{k -1}(> \tau) \rvert} \leq 1 + \frac{\mathbb{E} \lvert I_{k'}(> \tau) \rvert + \dots + \mathbb{E} \lvert I_{k-1}(> \tau) \rvert}{\mathbb{E}\lvert I_{k}(> \tau) \rvert}.\]
    Repeatedly applying \Cref{prop_batch_decrease} to the numerator gives:
    \[1 + \frac{\mathbb{E} \lvert I_{k'}(> \tau) \rvert + \dots +
        \mathbb{E} \lvert I_{k-1}(> \tau) \rvert}{\mathbb{E}\lvert
        I_{k}(> \tau) \rvert} \leq 1 + (2^{k-k'} + \dots + 2^1) \leq
      2^{k-k' + 1}. \qedhere\]
\end{proof}

To see the utility of \Cref{lem_machine_remain}, suppose $\mathbb{E} \lvert J_k (> \tau) \rvert = m$. Then roughly the lemma says in expectation, we lose at most half of our remaining machines between each batch. However, in the weighted free time the coefficient $n/2^k$ (corresponding to the number of jobs delayed by the current batch) also halves between each batch. Thus, although we are losing half of our machines, only half as many jobs are affected by this loss. 

First, we bound the expectation of $\lvert J_k(> \tau) \rvert$ when $\tau$ is sufficiently large (i.e. for all possible realizations of $\tau_k$.) The proof uses a Chernoff bound along with the definition of big jobs; see \Cref{sec_appendix_concentration} for proof.

\begin{restatable}{lem}{bigjobsmallexpectation}\label{lem_big_job_small_expectation}
    Let $m = \Omega(1)$ be sufficiently large. Then there exists a constant $c \geq 0$ such that for all batches $k$ and thresholds $\tau > 2 \mathbb{E} F^*(n-n/2^k)$, we have $\mathbb{E} \lvert J_k(> \tau) \rvert \leq m + c \sqrt{m}$.
\end{restatable}

Now, because $\mathbb{E} \lvert J_k(> \tau) \rvert = O(m)$, we can
bound the deviation of $\lvert J_k(> \tau) \rvert$ by
$\widetilde{O}(\sqrt{m})$ with high probability.

We define the notation
$\lvert J_k(> \tau) \rvert \stackrel{\pm \Delta}{\approx} \mathbb{E}
\lvert J_k(> \tau) \rvert$ to denote the event \[ \lvert\; \lvert J_k(>
  \tau) \rvert - \mathbb{E} \lvert J_k(> \tau) \rvert  \; \rvert \leq
  \Delta. \]
The proof of the next lemma is a Chernoff-union argument; see \Cref{sec_appendix_concentration} for proof.

\begin{restatable}{lem}{bigjobchernoff}\label{lem_big_job_chernoff}
    Let $\Delta = O(\sqrt{m} \log n)$ and $m = \Omega(1)$ be sufficiently large. Then with probability at least $1 - \frac{1}{\poly(n)}$, the following events hold:
    \begin{equation}\label{event_delta_big}
        \{\lvert J_k(> \tau) \rvert \stackrel{\pm \Delta}{\approx} \mathbb{E} \lvert J_k(> \tau) \rvert \quad \forall \text{ batches $k$ and thresholds $\tau > 2 \mathbb{E} F^*(n-n/2^k)$}\}.
    \end{equation}   
\end{restatable}

Combining \Cref{lem_machine_remain} and \Cref{lem_big_job_chernoff}, we can show the number of remaining machines is concentrated as well. Here we also need to bring in the per-realization properties of \algname to handle the case where concentration is not enough to ensure that we have at least one remaining machine. This is the main result of this section. Recall that we defined $\tau_k = 2 \max(\mathbb{E} F^*(n - n/2^k), F^*(n - n/2^k))$, so in particular $\tau_k \geq 2 \mathbb{E} F^*(n - n/2^k)$.

\begin{lem}\label{lem_remaining_mach_concentration}
    Suppose Event (\ref{event_delta_big}) holds. Then for all pairs of batches $k' \leq k$, we have $m - \lvert J_{k-1}( > \tau_k) \rvert \geq (3\Delta)^{-1} 2^{-(k-k'+1)} (m - \lvert J_{k' -1}(> \tau_k) \rvert)$, where $\Delta = O(\sqrt{m} \log n)$.
\end{lem}
\begin{proof}
    Consider fixed batches $k' \leq k$, and let $\mu_k = \mathbb{E} \lvert J_k(> \tau_k) \rvert$ and $\mu_{k'} = \mathbb{E} \lvert J_k'(> \tau_k) \rvert$. Note that $\tau_k \geq 2 \mathbb{E} F^*(n - n/2^k) \geq 2 \mathbb{E} F^*(n - n/2^{k'})$, so Event (\ref{event_delta_big}) gives $\lvert J_k(> \tau_k) \rvert \stackrel{\pm \Delta}{\approx} \mu_k$ and $\lvert J_{k'}(> \tau_k) \rvert \stackrel{\pm \Delta}{\approx} \mu_{k'}$. Further, we may choose $\Delta = O(\sqrt{m} \log n)$ large enough so that $\mu_k \leq m + \Delta$. Using these approximations with \Cref{lem_machine_remain} gives:
    \begin{align*}
        m - \lvert J_k(> \tau_k) \rvert &= m + \Delta - \lvert J_k(> \tau_k) \rvert - \Delta\\
        &\geq m + \Delta - \mu_k - 2\Delta\\
        &\geq 2^{-(k-k'+1)}(m + \Delta - \mu_{k'}) - 2\Delta\\
        &\geq 2^{-(k-k'+1)}(m - \lvert J_k(> \tau_k) \rvert) - 2\Delta.
    \end{align*}
    Finally, by \Cref{prop_big_threshold}, $m - \lvert J_k(> \tau_k) \rvert \geq 1$, so rearranging gives:
    \[3\Delta(m - \lvert J_k(> \tau_k)\rvert) \geq m - \lvert J_k(> \tau_k) \rvert + 2\Delta \geq 2^{-(k-k'+1)}(m + \lvert J_k(> \tau_k) \rvert).\]
\end{proof}

To summarize, we showed that up to a multiplicative $\widetilde{O}(\sqrt{m})$-factor, the number of unclogged machines with respect to threshold $\tau_k$ at worst halves in each batch up to $k$.

\subsection{Bounding small-in-the-past jobs}\label{subsec_small_past}

Recall that our goal is to bound $\sum_k n/2^k \cdot \frac{\Vol(I_k(\leq \tau_k))}{m - \lvert J_{k-1}(> \tau_k) \rvert}$. To this end, consider fixed $k$. Because $I_k \subset J_k^* = \cup_{k' \leq k} I_k^*$ (by \Cref{lem_batch_prefix}), we can write:
\[\frac{\Vol(I_k(\leq \tau_k))}{m - \lvert J_{k-1}(> \tau_k) \rvert} = \sum_{k' \leq k} \frac{\Vol(I_k \cap I_{k'}^*(\leq \tau_{k'}))}{m - \lvert J_{k-1}(> \tau_k) \rvert} + \sum_{k' \leq k} \frac{\Vol(I_k \cap I_{k'}^*(> \tau_{k'}, \leq \tau_k))}{m - \lvert J_{k-1}(> \tau_k) \rvert}.\]
Thus, we split $I_k$ depending on which batch $\Opt$ decided to schedule that job in. Further, we split $I_k \cap I_{k'}^*$ (i.e. the jobs our algorithm does in batch $k$ that $\Opt$ did in the past batch $k' \leq k$) into the jobs that are small-in-the-past (size at most $\tau_{k'}$) and big-in-the-past (size greater than $\tau_{k'}$ and at most $\tau_k$.)

The goal of this section is to bound the small-in-the-past jobs. This formalizes the idea that the rate at which we lose machines, guaranteed by \Cref{lem_remaining_mach_concentration}, is offset by the number of jobs $\Opt$ is delaying, captured by the exponentially decreasing weights $n/2^k$. More precisely, if $\opt$ decides to do a small job from $I_k$ in an earlier batch, say $I_{k'}^*$, then $\opt$ is averaging this small volume over at most a $\widetilde{O}(\sqrt{m}) \cdot 2^{k-k'}$-factor more unclogged machines. However, the weight of this term in $\opt$'s weighted free time increased by a $2^{k-k'}$-factor as well, corresponding to the number of jobs delayed by batch $k'$. Thus, up to a $\widetilde{O}(\sqrt{m})$-factor, there is no benefit to doing the small-in-the-past jobs any earlier. We show the following.

\begin{lem}\label{lem_small_past}
    Suppose Event (\ref{event_delta_big}) holds. Then the small-in-the-past jobs satisfy:
    \[\sum_k n/2^k \cdot \sum_{k' \leq k} \frac{\Vol(I_k \cap I_{k'}^*(\leq \tau_{k'}))}{m - \lvert J_{k-1}(> \tau_k) \rvert} = \widetilde{O}(\sqrt{m}) \cdot \sum_k n/2^k \cdot \tau_k.\]
\end{lem}
\begin{proof}
    Because there are $O(\log n)$ batches, it suffices to show for
    fixed $k$ and $k' \leq k$ that we have \[ \frac{\Vol(I_k \cap
        I_{k'}^*(\leq \tau_{k'}))}{m - \lvert J_{k-1}(> \tau_k)
        \rvert} = O(\Delta) \cdot 2^{k-k'} \tau_{k'}, \] where $\Delta = O(\sqrt{m} \log n)$. Summing over all $k$ and $k' \leq k$ would give the desired result.
    
    We upper bound the numerator using $I_k \cap I_{k'}^* \subset I_{k'}^*$ and apply \Cref{lem_remaining_mach_concentration} to the denominator. This gives:
    \[\frac{\Vol(I_k \cap I_{k'}^*(\leq \tau_{k'}))}{m - \lvert J_{k-1}(> \tau_k) \rvert} = O(\Delta) \cdot 2^{k - k'} \frac{\Vol(I_{k'}^*(\leq \tau_{k'}))}{m - \lvert J_{k'-1}(> \tau_k) \rvert} = O(\Delta) \cdot 2^{k - k'} \frac{\Vol(I_{k'}^*(\leq \tau_{k'}))}{m - \lvert J_{k'-1}^*(> \tau_{k'}) \rvert}.\]
    In the final step, we used $J_{k' - 1} \subset J_{k' -1}^*$ (by \Cref{lem_batch_prefix}) and $\tau_k \geq \tau_{k'}$.
    
    Finally, we show \[ \frac{\Vol(I_{k'}^*(\leq \tau_{k'}))}{m -
        \lvert J_{k'-1}^*(> \tau_{k'}) \rvert} = O(\tau_{k'}) .\]
    Recall that $\tau_{k'} > F^*(n-n/2^{k'})$, so $\Opt$ schedules at
    most one $\tau_{k'}$-big job per machine in $J_{k'}^*$. Further,
    $\Opt$ schedules $I_{k'}^*(\leq \tau_{k'})$ only on the $m -
    \lvert J_{k' - 1}^*(> \tau_{k'}) \rvert$ machines that have not
    yet scheduled a $\tau_{k'}$-big job yet. By averaging, after
    scheduling $I_{k'}^*(\leq \tau_{k'})$, every such machine in
    $\Opt$ has load at least \[ \frac{\Vol(I_{k'}^*(\leq
        \tau_{k'}))}{m - \lvert J_{k'-1}^*(> \tau_{k'}) \rvert} -
      \tau_{k'}.\] One of these machines must achieve the free time
    $F^*(n-n/2^{k'})$, because every other machine has already
    scheduled a $\tau_{k'}$-big job. This implies \[ F^*(n-n/2^{k'})
      \geq \frac{\Vol(I_{k'}^*(\leq \tau_{k'}))}{m - \lvert
        J_{k'-1}^*(> \tau_{k'}) \rvert} - \tau_{k'}. \] Rearranging and using $\tau_{k'} > F^*(n-n/2^{k'})$ give the desired result.
\end{proof}

Thus, the contribution of the small-in-the-past jobs to $\Alg$'s weighted free time is comparable to $\Opt$'s weighted free time,  up to a $\widetilde{O}(\sqrt{m})$-factor.

\subsection{Bounding big-in-the-past jobs}\label{subsec_big_past}

The goal of this section is to bound the big-in-the-past jobs, that is:
\[\sum_k n/2^k \cdot \sum_{k' \leq k} \frac{\Vol(I_k \cap I_{k'}^*(> \tau_{k'}, \leq \tau_k))}{m - \lvert J_{k-1}(> \tau_k) \rvert}.\]
For convenience, we define $I_{kk'} = I_k \cap I_{k'}^*(> \tau_{k'}, \leq \tau_k)$. Note that we cannot apply volume arguments as in \S \ref{subsec_small_past}, because the big-in-the-past jobs are $\tau_{k'}$-big. Instead, we will use the fact that $\Opt$ schedules at most one $I_{kk'}$-job per machine.

There are two types of jobs in $j \in I_{kk'}$: We say $j$ is \emph{blocked} if $\Opt$ later schedules a $\tau_k$-big job in $J_{k-1}$ on the same machine as $j$ (recall that $J_{k-1} \subset J_{k-1}^*$ by \Cref{lem_batch_prefix}.) Otherwise, $j$ is \emph{unblocked}. Further, a machine is blocked/unblocked if the $I_{kk'}$-job scheduled on that machine is blocked/unblocked. Thus we can partition $I_{kk'} = B_{kk'} \cup U_{kk'}$ into blocked and unblocked jobs, respectively.

By splitting the volume of jobs into unblocked and blocked, we can rewrite:
\[\sum_k n/2^k \cdot \sum_{k' \leq k} \frac{\Vol(I_k \cap I_{k'}^*(> \tau_{k'}, \leq \tau_k))}{m - \lvert J_{k-1}(> \tau_k) \rvert}  = \sum_k n/2^k \cdot \sum_{k' \leq k} \frac{\Vol(U_{kk'})}{m - \lvert J_{k-1}(> \tau_k) \rvert} + \sum_k n/2^k \cdot \frac{\Vol(B_{kk'})}{m - \lvert J_{k-1}(> \tau_k) \rvert}.\]
Intuitively, the unblocked jobs are not problematic because there can be at most $m - \lvert J_{k-1}(> \tau_k) \rvert$ such jobs.

\begin{lem}\label{lem_unblocked}
    The unblocked jobs satisfy $\sum_k n/2^k \cdot \sum_{k' \leq k} \frac{\Vol(U_{kk'})}{m - \lvert J_{k-1}(> \tau_k) \rvert} \leq O(\log n) \cdot \sum_k n/2^k \cdot \tau_k$.   
\end{lem}
\begin{proof}
    Because there are $O(\log n)$ batches, it suffices to show for
    fixed $k$ and $k' \leq k$ that \[ \frac{\Vol(U_{kk'})}{m - \lvert J_{k-1}(> \tau_k) \rvert} \leq \tau_k.\] We recall that every job in $U_{kk'}$ is $\tau_k$-small, so:
    \[\frac{\Vol(U_{kk'})}{m - \lvert J_{k-1}(> \tau_k) \rvert} \leq \tau_k \cdot \frac{\lvert U_{kk'} \rvert}{m - \lvert J_{k-1}(> \tau_k) \rvert}.\]
    
    We note that every job in $U_{kk'}$ is $\tau_{k'}$-big, and $\Opt$ schedules these jobs in batch $I_{k'}^*$. Thus, there is at most one $U_{kk'}$-job per unblocked machine. Further, there are at most $m - \lvert J_{k-1}(> \tau_k) \rvert$ unblocked machines, because each $\tau_k$-big job in $J_{k-1}$ must be scheduled on a separate machine of $\Opt$ (because $J_{k-1} \subset J_k^*$ by \Cref{lem_batch_prefix}.) We conclude, $\frac{\lvert U_{kk'} \rvert}{m - \lvert J^*_{k-1}(> \tau_k) \rvert} \leq 1$, as required.
\end{proof}

It remains to handle the blocked jobs. Again, the central issue is that $\Opt$ does blocked jobs in an earlier batch before some machines get clogged. On the other hand, \choosealg puts these jobs in a later batch when we have fewer machines.

Unlike our previous arguments, for the blocked jobs we will charge the volume of these jobs to the completion time of $\Opt$ directly. Because these jobs are blocked, $\Opt$ must schedule a $\tau_k$-big job later on the same machine. In particular, $\Opt$ must have kept scheduling Bernoulli jobs with size parameter at least $\tau_k$ until one comes up heads. We will charge $B_{kk'}$ to the completion time of all of these coin flips.

As before, we consider a fixed threshold $\tau$ and later union bound over all relevant thresholds. In this section, for any batch $k$ and threshold $\tau$, we define $p_{k \tau} \in [0,1]$ to be the largest probability parameter across all jobs $j$ in $J_{k-1}$ with $s_j > \tau$ (if no such job exists, then we follow the convention $p_{k \tau} = 0$.) Note that $p_{k \tau}$ is deterministic for fixed $\tau$. We first relate the number of remaining machines with the expected number of heads in the $k$th batch.

\begin{prop}\label{prop_blocked_num_delayed}
    Consider any batches $k$, $k' \leq k$ and threshold $\tau \geq 2 \mathbb{E} F^*(n-n/2^k)$. Suppose Event (\ref{event_delta_big}) holds. Then we have $m- \lvert J_{k-1}(> \tau) \rvert \geq p_{k \tau} \cdot n/2^k - O(\Delta)$, where $\Delta = O(\sqrt{m} \log n)$.
\end{prop}
\begin{proof}
    First, if $p_{k \tau} = 0$, then $\lvert J_{k-1}(> \tau) \rvert = 0$, so the proposition is trivial. Thus, we may assume $p_{k \tau} > 0$. In particular, \choosealg included at least one job $j \in J_{k-1}$ with $s_j > \tau$ and $p_j = p_{k \tau}$. It follows, \choosealg will include $n/2^k$ further jobs in $I_k$ with size parameter larger than $\tau_k$ and probability parameter at least $p_{k \tau}$. Thus, we have $\mathbb{E} \lvert I_k(> \tau) \rvert \geq p_{k \tau} \cdot n/2^k$. Rewriting $\mathbb{E} \lvert I_k(> \tau) \rvert = \mathbb{E} \lvert J_k(> \tau) \rvert - \mathbb{E} \lvert J_{k-1}(> \tau) \rvert$ and applying \Cref{lem_big_job_small_expectation} and Event (\ref{event_delta_big}) to the first and second expectations, respectively gives:
    \[p_{k \tau} \cdot n/2^k \leq \mathbb{E} \lvert I_k(> \tau) \rvert = \mathbb{E} \lvert J_k(> \tau) \rvert - \mathbb{E} \lvert J_{k-1}(> \tau) \rvert \leq (m + O(\sqrt{m})) - (\lvert J_{k-1}(> \tau) \rvert - O(\Delta)).\]
    Rearranging gives the desired result.
\end{proof}

To see the utility of~\Cref{prop_blocked_num_delayed}, we assume for a moment that $\tau_k$ is deterministic and ignore the additive $O(\Delta)$ term in the proposition. Then we could rewrite $n/2^k \cdot \frac{\Vol(B_{kk'})}{m - \lvert J_{k-1}(> \tau_k) \rvert} \lesssim \frac{1}{p_{k \tau}} \cdot \Vol(B_{kk'})$.

To relate this expression with $\Opt$, we note that $\Opt$ schedules a $\tau_k$-big job on top of each job in $B_{kk'}$. In particular, $\Opt$ must schedule enough Bernoulli jobs $j$ with $s_j > \tau_k$ until at one comes up heads on each such machine. Each such job also satisfies $p_j \leq p_{k \tau_k}$, so - roughly - in $\Opt$ we expect each blocked job to delay at least $\frac{1}{p_{k \tau}}$ jobs in order for that machine to become blocked. This would give $\frac{1}{p_{k \tau_k}} \cdot \Vol(B_{kk'}) \lesssim \Opt$, as required.

\subsection{Coin Game}\label{subsec_coin}

It remains to formalize this idea using a martingale argument.  We begin by defining an (artificial) game, which will model the process of a machine becoming blocked.

\begin{definition}[Coin Game]\label{def_coin_game}
    The game is played with $n$ coins and $m$ machines by a single player. The coins are independent such that coin $j$ comes up heads with probability $p_j$. Initially, all machines are \emph{available}.
      At each turn, the player can either choose to flip a previously unflipped coin on an available machine or to end the game. In the former case, if the coin comes up heads, then the machine becomes unavailable.
    The game ends when the player chooses to, or if we run out of unflipped coins or available machines.
\end{definition}

Now we are ready to interpret $\Opt$ as implicitly playing a coin game to block machines.

\begin{definition}[Induced Coin Game]\label{def_induced_coin_game}
    Consider pairs of batches $k' \leq k$ and thresholds $\tau' \leq \tau$. Then the $(k', k, \tau', \tau)$-induced coin game (with respect to policy $\Opt$) is a distribution over coin games defined as follows:
    \begin{OneLiners}
        \item The machines are the ones of $\Opt$ whose final job in $J_{k'}^*$ has size exactly $\tau'$.
        \item For every job in $j \in J_{k-1} \setminus J_{k'}^*$ with $s_j > \tau$, we have a coin with the same probability parameter.
        \end{OneLiners}
        The player of the coin game simulates $\Opt$ as follows. Starting from after $\Opt$ schedules $J_{k'}^*$, if $\Opt$ subsequently schedules a job on a machine that is still available (in the coin game), then the player flips the corresponding coin (if such a coin exists) on the same machine. The player decides to stop when it runs of out coins or all machines are unavailable.
\end{definition}

One should imagine that the machines in the induced coin game are exactly those that can become blocked. Thus, a machine becoming unavailable in the coin game corresponds to it becoming blocked in $\Opt$, and the total number of flipped coins records how many jobs were delayed by $\tau'$.

Using a martingale argument, we relate the number of machines that become unavailable with the number of flipped coins. The next lemma formalizes the idea that to block a machine, we expect $\Opt$ to flip $\frac{1}{p_{k\tau_k}}$ coins per blocked machine. Recall that for any batch $k$ and threshold $\tau$, we define $p_{k\tau}$ to be the largest probability parameter across all jobs $j$ in $J_{k-1}$ with $s_j > \tau$.

\begin{lem}\label{lem_induced_coin_game_concentration}
    With probability $1 - \frac{1}{\poly(n)}$, the following event holds:
    \begin{equation}\label{event_coin_cost}
        \{\#(\text{unavailable machines}) \leq p_{k\tau} \cdot \#(\text{flipped coins}) + \Delta \quad  \text{$\forall$ $(k', k, \tau', \tau)$- induced coin games}\},
    \end{equation}   
    where $\Delta = O(\sqrt{m} \log n)$.
\end{lem}
\begin{proof}
    Because there are $O(\log n)$ batches and $L = O(\log n)$ relevant thresholds, by union-bounding over all pairs of batches and thresholds, it suffices to show that a fixed $(k',k,\tau',\tau)$-induced coin game satisfies:
    \[\mathbb{P}(\#(\text{unavailable machines}) \leq p_{k\tau} \cdot \#(\text{flipped coins}) + \Delta) = 1 - \frac{1}{\poly(n)}.\]
    We will define a martingale to count the number of unavailable machines. For all $t \geq 0$, let $A_t$ be the (adaptively chosen) set of the first $t$ coins flipped by the player. If the player stops before flipping $t$ coins, then we define $A_t = A_{t-1}$. Now consider the sequence of random variables $M_t = \sum_{j \in A_t} C_j - \sum_{j \in A_t} p_j$ for all $t \geq 0$, where $C_j \sim Ber(p_j)$ is the distribution of coin $j$. Note that $\sum_{j \in A_t} C_j$ is exactly the number of heads in the first $t$ coin flips, which is the number of unavaiable machines.
    
    We claim that $M_t$ is a martingale. Consider any $t \geq 0$. There are two cases. If $A_t = A_{t-1}$, then $M_t = M_{t-1}$, so trivially $\mathbb{E}[M_t \mid M_{t-1}, \dots, M_0] = M_{t-1}$. Otherwise, $A_t = A_{t-1} \cup \{j\}$ for some adaptively chosen coin $j$. It suffices to show the martingale property conditioned on the next coin being $j$ for any fixed coin $j$:
    \begin{align*}
       \mathbb{E}[M_t \mid M_{t-1}, \dots, M_0, A_t = A_{t-1} \cup \{j\}] &= \mathbb{E}[M_{t-1} + C_j - p_j\mid M_{t-1}, \dots, M_0, A_t = A_{t-1} \cup \{j\}]\\
       &= M_{t-1} + p_j - p_j = M_{t-1},
    \end{align*}
    as required.
    
    To bound the deviation of $M_t$, we apply Freedman's inequality~\cite{10.1214/aop/1176996452} to the martingale difference sequence of $M_t$.

    \begin{prop}[Freedman's inequality]\label{prop_freedman} 
        Consider a real-valued martingale sequence $\{X_t\}_{t \geq 0}$ such that $X_0 = 0$ and $\lvert X_t \rvert \leq M$ almost surely for all $t$. Let $Y_t = \sum_{s = 0}^t \mathbb{E}[X_s^2 \mid X_{s -1}, \dots, X_0]$ denote the quadratic variation process of $\{X_t\}_t$. Then for any $\ell \geq 0, \sigma^2 > 0$ and stopping time $\tau$, we have:
        \[\mathbb{P}(\lvert \sum_{t = 0}^\tau X_t \rvert \geq \ell \text{ and } Y_\tau \leq \sigma^2) \leq 2 \cdot \exp\bigg(- \frac{\ell^2 / 2}{\sigma^2 + M \ell / 3}\bigg).\]
    \end{prop}

    We let $X_t$ denote the martingale difference sequence of $M_t$, which is defined as $X_0 = 0$ and $X_t = M_t - M_{t-1}$ for all $t > 0$. Because $M_t$ is a martingale, $X_t$ is as well. Furthermore, we have $\lvert X_t \rvert \leq  1$ almost surely for all $t$. For any $t \geq 0$, we let $j_t$ be the (adaptively chosen) $t$th coin flip. Then we can bound the quadratic variation process by:
    \begin{align*}
        Y_t = \sum_{s = 0}^t \mathbb{E}[X_s^2 \mid X_{s -1}, \dots, X_0] &= \sum_{s = 0}^t \mathbb{E}[(C_{j_s} - p_{j_s})^2 \mid X_{s-1}, \dots, X_0]\\
        &\leq \sum_{s = 0}^t \mathbb{E}[C_{j_s}^2 \mid X_{s-1}, \dots, X_0]\\
        &= \sum_{s = 0}^t \mathbb{E}[C_{j_s} \mid X_{s-1}, \dots, X_0].
    \end{align*}
    Note that the $C_{j_1} + \dots C_{j_t} \leq m$ almost surely, because the induced coin game has at most $m$ machines, and any adaptive policy can flip at most one heads per machine. Thus, we have $Y_t \leq m$ for all $t$.
    
    Now let $T$ be the stopping time when the induced coin game ends, so $T$ is exactly the number of flipped coins. Then Freedman's inequality gives:
    \[\mathbb{P}(\lvert \sum_{t = 0}^T X_t \rvert \geq \Delta) = \mathbb{P}(\lvert \sum_{t = 0}^T X_t \rvert \geq \Delta \text{ and } Y_T \leq m) \leq 2 \cdot \exp(- \frac{\Delta^2/2}{m + \Delta/3}).\]
    Taking $\Delta = O(\sqrt{m} \log n)$ gives $\mathbb{P}(\lvert \sum_{t = 0}^T X_t \rvert \geq \Delta) \leq \frac{1}{\poly(n)}$.
    
    Finally, we observe that $\#(\text{unavailable machines}) = \sum_{j \in A_T} C_j$. Further, we have $p_{k\tau} \cdot \#(\text{flipped coins}) \geq \sum_{j \in A_T} p_j$, because every coin $j$ corresponds to a job in $J_{k-1}$ with $s_j > \tau$, so $p_j \leq p_{k\tau}$ for all coins. Thus, we conclude:
    \[\mathbb{P}(\#(\text{unavailable machines}) > p_{k \tau} \cdot \#(\text{flipped coins}) + \Delta) \leq \mathbb{P}(\lvert \sum_{t = 0}^T X_t \rvert > \Delta) \leq \frac{1}{\poly(n)}.\]
\end{proof}

Combining \Cref{prop_blocked_num_delayed} and \Cref{lem_induced_coin_game_concentration}, we can bound the blocked jobs:

\begin{lem}\label{lem_blocked}
    Suppose Events (\ref{event_delta_big}) and (\ref{event_coin_cost}) hold. Then the blocked jobs satisfy:
    \[\sum_k n/2^k \cdot \sum_{k' \leq k} \frac{\Vol(B_{kk'})}{m - \lvert J_{k-1}(> \tau_k) \rvert} = \widetilde{O}(\sqrt{m}) (\sum_k n/2^k \cdot \tau_k + \Opt).\]
\end{lem}
\begin{proof}
    Because there are $O(\log n)$ batches $k$, it suffices to show for fixed $k, k' \leq k$ that $n/2^k \cdot \frac{\Vol(B_{kk'})}{m - \lvert J_{k-1}(> \tau_k) \rvert} = O(\Delta \log n) ( n/2^k \cdot \tau_k + \Opt)$ for $\Delta = O(\sqrt{m} \log n)$. We consider two cases.
    
    First, on the event that $p_{k\tau_k} = 0$, we have $m - \lvert J_{k-1}(> \tau_k) \rvert = m$. Recall that every job in $B_{kk'}$ is $\tau_{k'}$-big and in $I_{k'}^*$, so there is at most one such job per machine in $\Opt$. Then we can bound:
    \[n/2^k \cdot \frac{\Vol(B_{kk'})}{m - \lvert J_{k-1}(> \tau_k) \rvert} \leq n/2^k \cdot \tau_k \frac{m}{m} = n/2^k \cdot \tau_k.\]
    
    Otherwise, we have $p_{k\tau_k} > 0$. Here we related the blocked jobs to the induced coin games:
    \begin{align*}
        \Vol(B_{kk'}) &= \sum_{\tau' \leq \tau_k} \tau' \cdot \lvert \{j \in B_{kk'} \mid X_j = \tau'\} \rvert\\
        &= \sum_{\tau' \leq \tau_k} \tau' \cdot \#(\text{unavailable machines in $(k',k, \tau', \tau_k)$-induced coin game})\\
        &\leq \sum_{\tau' \leq \tau_k} \tau' \cdot (p_{k\tau_k} \cdot \#(\text{flipped coins in $(k',k, \tau', \tau_k)$-induced coin game}) + \Delta)\\
        &\leq O(\log n) \cdot p_{k \tau_k} \cdot \Opt + O(\Delta \log n) \cdot \tau_k.
    \end{align*}
    where the first inequality follows from Event (\ref{event_coin_cost}). The second follows because there are $O(\log n)$ relevant thresholds $\tau' \leq \tau_k$, and every flipped coin in the $(k',k, \tau', \tau_k)$-induced coin game corresponds to $\Opt$ scheduling a job on a machine that already scheduled some job with size $\tau'$, so every such job has completion time at least $\tau'$. It follows:
    \[n/2^k \cdot \frac{\Vol(B_{kk'})}{m - \lvert J_{k-1}(> \tau_k) \rvert} \leq n/2^k \cdot O(\log n) \frac{p_{k\tau_k}}{m - \lvert J_{k-1}(> \tau_k) \rvert} \cdot \Opt + n/2^k \cdot O(\Delta \log n) \frac{\tau_k}{m - \lvert J_{k-1} \rvert}.\]
    By \Cref{prop_blocked_num_delayed}, we can bound the first term by:
    \begin{align*}
    n/2^k \cdot O(\log n) \frac{p_{k\tau_k}}{m - \lvert J_{k-1}(>
      \tau_k) \rvert} \cdot \Opt &= O(\log n) \frac{m - \lvert
                                   J_{k-1}(> \tau_k) \rvert +
                                   O(\Delta)}{m - \lvert J_{k-1}(>
                                   \tau_k) \rvert} \cdot \Opt \\ &= O(\Delta \log n) \cdot \Opt.
    \end{align*}
    We can bound the second term by:
    \[n/2^k \cdot O(\Delta \log n) \frac{\tau_k}{m - \lvert J_{k-1}(> \tau_k) \rvert} = O(\Delta \log n) \cdot n/2^k \cdot \tau_k.\]
    Combining both bounds completes the proof.
\end{proof}

We summarize our bounds for the unblocked and blocked jobs by the next lemma, which follows immediately from \Cref{lem_unblocked} and \Cref{lem_blocked}.

\begin{lem}\label{lem_blocked_and_unblocked}
    Suppose Events (\ref{event_delta_big}) and (\ref{event_coin_cost}) hold. Then the big-in-the-past jobs satisfy:
    \[\sum_k n/2^k \cdot \sum_{k' \leq k} \frac{\Vol(I_k \cap I_{k'}^*(> \tau_{k'}, \leq \tau_k))}{m - \lvert J_{k-1}(> \tau_k) \rvert} = \widetilde{O}(\sqrt{m})\cdot \bigg(\sum_k n/2^k \cdot \tau_k + \Opt\bigg).\]
\end{lem}

\subsection{Putting it all together}\label{subsec_put_together}

We are ready to prove \Cref{thm_main_alg} and \Cref{thm_free_main}, which we restate here for convenience.

\mainalg*

\freemain*

\Cref{thm_free_main} follows from partitioning $\alg$'s weighted free time into the contribution due to small-in-the-past and big-in-the-past jobs (which we further partitioned into unblocked and blocked jobs.)

\begin{proof}[Proof of \Cref{thm_free_main}]

    We assume $m = \Omega(1)$ is sufficiently large. Then we complete the proof of \Cref{thm_free_main} by combining our bounds for the small-in-the-past- and big-in-the-past jobs. We bound $\Alg$'s weighted free time by \Cref{prop_k_free_vol}:
\[\sum_k n/2^k \cdot F(J_k) = O (\sum_k n/2^k \cdot \frac{\Vol(I_k(\leq \tau_k))}{m- \lvert J_{k-1}(> \tau_k) \rvert} + \sum_k n/2^k \cdot \tau_k).\]
We recall $\tau_k = 2 \cdot \max(\mathbb{E} F^*(n - n/2^k), F^*(n - n/2^k))$, so $\mathbb{E} \tau_k = O(\mathbb{E} F^*(n-n/2^k))$. Thus, in expectation, the second sum is at most:
\[\mathbb{E} \sum_k n/2^k \cdot \tau_k = O(\mathbb{E} \sum_k n/2^k \cdot F^*(n-n/2^k)).\]

It remains to bound the first sum, which we split into the contribution due to small-in-the-past and big-in-the-past jobs:
\[\sum_k n/2^k \cdot \frac{\Vol(I_k(\leq \tau_k))}{m - \lvert J_{k-1}(> \tau_k) \rvert} = \sum_k n/2^k \cdot \sum_{k' \leq k} \frac{\Vol(I_k \cap I_{k'}^*(\leq \tau_{k'}))}{m - \lvert J_{k-1}(> \tau_k) \rvert} + \sum_k n/2^k \cdot \sum_{k' \leq k} \frac{\Vol(I_k \cap I_{k'}^*(> \tau_{k'}, \leq \tau_k))}{m - \lvert J_{k-1}(> \tau_k) \rvert}.\]
On Events (\ref{event_delta_big}) and (\ref{event_coin_cost}), we can apply \Cref{lem_small_past} to the first term and \Cref{lem_blocked_and_unblocked} to the second to obtain:
\[\sum_k n/2^k \cdot \frac{\Vol(I_k(\leq \tau_k))}{m - \lvert J_{k-1}(> \tau_k) \rvert} = \widetilde{O}(\sqrt{m}) (\sum_k n/2^k \cdot \tau_k + \Opt).\]
Again, in expectation, this contributes $\widetilde{O}(\sqrt{m}) (\mathbb{E} \sum_k n/2^k \cdot F^*(n-n/2^k) + \mathbb{E} \Opt)$ to $\Alg$'s expected weighted free time.

Finally, we consider the event that Event (\ref{event_delta_big}) or
Event (\ref{event_coin_cost}) does not hold. Recall that by
\Cref{lem_big_job_chernoff} and
\Cref{lem_induced_coin_game_concentration}, this happens with
probability at most $\frac{1}{\poly(n)}$ because $m = \Omega(1)$ is sufficiently large. Further, on this event, we
can trivially upper bound $\sum_k n/2^k \cdot \frac{\Vol(I_k(\leq
  \tau_k))}{m - \lvert J_{k-1}(> \tau_k) \rvert} = \poly(n)$, because
there are $n$ jobs each with size at most $\poly(n)$ almost
surely. Thus, the contribution of this event to the overall
expectation is $O(1)$. We
conclude, $\Alg$'s expected weighted free time is at most
$\widetilde{O}(\sqrt{m}) (\mathbb{E} \sum_k n/2^k \cdot F^*(n-n/2^k) +
\mathbb{E} \Opt) + O(1)$.
\end{proof}

To complete the proof of \Cref{thm_free_main}, we relate the weighted free time to the completion time. We also use our warm-up $\tilde{O}(m)$-approximation when $m$ is too small to apply \Cref{thm_main_alg}.

\begin{proof}[Proof of \Cref{thm_main_alg}]
    First, suppose $m = \Omega(1)$ is sufficiently large. Then by \Cref{thm_free_main}, $\alg$'s weighted free time satisfies:
    \[\mathbb{E}\bigg[ \sum_k n/2^k \cdot F(J_k)\bigg] =
      \widetilde{O}(\sqrt{m}) \cdot \bigg( \mathbb{E} \bigg[\sum_k n/2^k \cdot
      F^*(n-n/2^k)\bigg] + \mathbb{E}[\Opt\,] \bigg) + O(1).\]
    Applying \Cref{lem_new_obj} to relate weighted free time to completion time gives:
    \[\mathbb{E} [\alg] = \widetilde{O}(1) \cdot \mathbb{E}\bigg[ \sum_k n/2^k \cdot F(J_k)\bigg] + \widetilde{O}(\mathbb{E}[\opt]) = \widetilde{O}(\sqrt{m}) \cdot \bigg(\mathbb{E}[\opt] + O(1)\bigg).\]
    This gives the desired guarantee if $m = \Omega(1)$.
    
    Otherwise, if $m = O(1)$, \Cref{lem_warm_up_m} immediately gives that \algname is $\widetilde{O}(m) = \widetilde{O}(1)$-approximate, so $\mathbb{E} [\alg] = \widetilde{O}(\mathbb{E}[\opt])$.
\end{proof}

This completes the analysis of \algname. Since the proof had several
conceptual parts, let us a give a quick summary.

\paragraph{Summary.}
Recall that our analysis began by passing from completion time to our
new proxy objective: weighted free time in
\S\ref{subsec_free_time}. As we mentioned earlier, a key benefit of
working with free times rather than completion times was that we could
completely control what jobs we \emph{started} to achieve the $i$th
free time, whereas we have far less control over the first $i$ jobs to \emph{finish}. This allowed us to make the contribution of each job to the weighted free time more modular: either the job contributed to the small volume in a batch, or it contributed to the clogged machines---see \Cref{eq_free_time_vol}. We then controlled the rate at which $\alg$ and $\opt$ clog up machines in \S\ref{subsec_bound_clog}. Then in \S\ref{subsec_small_past}-\ref{subsec_coin} we compared the times at which $\alg$ and $\opt$ chose to do the same volume of small jobs. Since these were the only two ways in which a job affected the weighted free time, we could combine these two ideas in \S\ref{subsec_put_together} to complete our analysis.









\section{Conclusion}\label{sec_conclusion}

We gave an improved approximation for stochastic completion times,
which does not depend on the job size variances, and has a sublinear
dependence on the number of machines $m$. Observe that the weighted
free time is a valid proxy objective for \emph{any} job size
distributions, not just Bernoulli jobs, so extending our result to
general stochastic jobs requires us to solve subset selection and
batch free time minimization for these settings. 

Many interesting open problems remain: can we improve our
approximation ratio even further? We also do not have a good grasp on
the complexity of this problem: is the stochastic problem provably
hard to solve/approximate? Can we use the idea of passing from
completion times to free times more broadly? Can we extend the results
to other scheduling objectives, such as flow/response times? In general, stochastic scheduling problems (apart from the makespan objective) are not well understood from a distribution-independent approximation perspective, and we hope that our work will lead to further interesting developments.

{\small
\bibliographystyle{alpha}
\bibliography{ref}
}

\appendix

\section{Sensitivity of number of machines}\label{sec_appendix_num_machines}

For a fixed collection of jobs, and any number of machines $m$, we let $\opt(m)$ be the optimal completion time for these jobs on $m$ machines.

\begin{lem}\label{lem_num_mach_stoch}
    For any number of machines $m$ sufficiently large, there exists a collection of \emph{identical} Bernoulli jobs with $\mathbb{E} \opt(\frac{m}{2}) = e^{\Omega(m)} \cdot \mathbb{E} \opt(m)$.
\end{lem}
\begin{proof}
    We fix a number of machines $m$. Define $L = e^{cm}$ for a constant $c > 0$. Then consider the collection of $\frac{7}{8} mL$ Bernoulli jobs distributed as $\Ber(\frac{1}{L})$. Note that because jobs are identically distributed, we may assume $\opt$ list schedules jobs in arbitrary order.
    
    We first claim that $\mathbb{E} \opt(m) = O(m)$. To see this, let $H \sim \Binom(\frac{7}{8}mL, \frac{1}{L})$ be the number of jobs that come up heads. On the event $H \leq m$, each machine schedules some number of jobs with realized size zero and then at most one job with realized size $1$. Thus, on this event we have $\opt(m) \leq H$. Further, by Chernoff (\Cref{prop_chernoff}), we have:
    \[\mathbb{P}(H > m) \leq \mathbb{P}(H \geq \mathbb{E}[H] + \frac{m}{8}) = e^{-\Theta(m)}.\]
    We conclude, the contribution of the event $H \leq m$ to $\mathbb{E} \opt(m)$ is at most $\mathbb{E} H = \frac{7}{8}m$, and the contribution of the event $H < m$ is at most $\poly(mL) \cdot \mathbb{P}(H > m) = \poly(m e^{cm}) \cdot e^{-\Theta(m)} = O(1)$ for $c$ sufficiently small. This gives $\mathbb{E} \opt(m) \leq \frac{7}{8} m + O(1) = O(m)$.
    
    On the other hand, we have $\mathbb{E} \opt(\frac{m}{2}) = \Omega(mL)$. Let $H' \sim \Binom(\frac{3}{4}mL, \frac{1}{L})$ be the number of heads among the first $\frac{3}{4}mL$ jobs. Analogously by Chernoff we have $\mathbb{P}(H' < \frac{m}{2}) \leq e^{-\Theta(m)}$. Thus, on the event $H' \geq \frac{m}{2}$ (which happens with probability $(1 - o(1))$, after scheduling the first $\frac{3}{4}mL$ jobs, each of the $\frac{m}{2}$ machines has a job of realized size $1$. It follows, the remaining $\Omega(mL)$ unscheduled jobs all have completion times at least $1$. Thus, we can lower bound $\mathbb{E} \opt(\frac{m}{2}) = \Omega(mL) \cdot (1 - o(1))$. Taking $m$ sufficiently large gives the desired gap.
\end{proof}

\begin{lem}\label{lem_num_mach_det}
    For any collection of deterministic jobs and any $m \geq 2$, we have $\opt(\frac{m}{2}) \leq 3 \cdot \opt(m)$.
\end{lem}
\begin{proof}
    Consider the schedule achieving $\opt(m)$, and let $C_j^m$ be the completion time of job $j$ in this schedule. We construct a schedule on $\frac{m}{2}$ machines with completion time at most $3 \cdot \opt(m)$. Our algorithm is to list schedule the jobs on $\frac{m}{2}$ machines in increasing order of $C_j^m$.
    
    Let $C_j$ be the completion time of job $j$ in this schedule. We claim that $C_j \leq 3 C_j^m$ for all jobs $j$, which gives the desired result. Assume for contradiction that this is not the case, so let $j$ be the first job with $C_j > 3 C_j^m$. It must be the case that up until time $2C_j^m$, all $\frac{m}{2}$ machines are busy running jobs $j'$ with $C_{j'}^m \leq C_j^m$. The total size of such $j'$ jobs is strictly larger than $\frac{m}{2} \cdot 2 C_j^m = m \cdot C_j^m$. However, $\opt(m)$ must complete all such $j'$ jobs by time $C_j^m$. This is a contradiction.
\end{proof}

\section{Exchange Argument}\label{sec_appendix_exchange}

\exchange*
\begin{proof}
    We suppose there exist jobs $a,b$ such that $X_a \sim s \cdot Ber(p_a)$ and $X_b \sim s \cdot Ber(p_b)$ with $p_a \leq p_b$ such that the optimal completion time policy schedules $b$ before $a$ in some realization. Consider the decision tree corresponding to this policy (described in \S \ref{sec_subset_selection}.) Thus, we assume this tree schedules $b$ before $a$ is some realization (i.e. some root-leaf path.) It follows, there exists a subtree rooted at $b$ such that $a$ is scheduled on each root-leaf path of this subtree. We denote this subtree by $T$. Entering this subtree, the machines have some fixed initial loads and $T$ schedules a fixed set of jobs $J$.
    
    We will modify the subtree $T$ so that we start $a$ before $b$ on each root-leaf path. Further, this will not increase the expected completion time of the overall schedule. We construct the modified subtree $T'$ as follows. Let the left- and right subtrees (corresponding to the root job $b$ coming up size $0$ or $s$) of $T$ be $T_L$ and $T_R$, respectively. $T'$ is rooted at job $a$. In $T'$, the right subtree of $a$ is $T_R$, but with job $a$ replaced by job $b$. We denote this modifed subtree by $T_R(a \rightarrow b)$. On the left subtree of $a$, first, independently of all jobs, we flip a coin that comes up heads with probability $q$. We will choose $q$ later. If the coin is tails, then we schedule subtree $T_L$ with job $a$ replaced by job $b$, so $T_L(a \rightarrow b)$. Otherwise, if the coin is heads, then we schedule $b$. The left- and right subtrees of $b$ are $T_L$ and $T_R$, except the job $a$ is replaced by a dummy job that is always zero. In particular, this job does not contribute to the completion time, but upon reaching this node we will always follows the left subtree. We denote these subtrees by $T_L(-a), T_R(-a)$, respectively. This completes the description of $T'$. See
    Figure~\ref{fig:exch_appendix} for the modified tree $T'$.
    
    \begin{figure}
        \centering
        \includegraphics[width=0.75\textwidth]{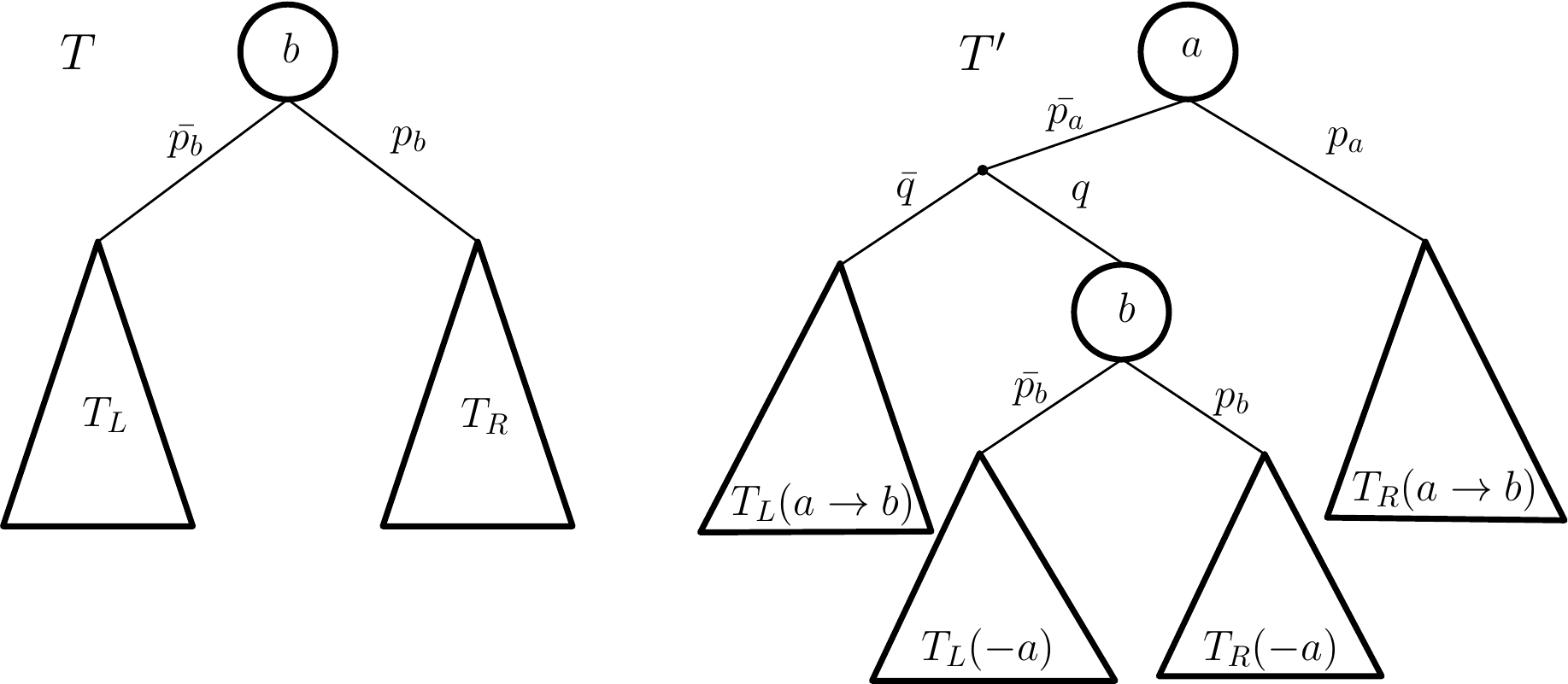}
        \caption{Original and modified decision trees}
        \label{fig:exch_appendix}
    \end{figure}
    
    Note that $T'$ schedules the same jobs as $T$ and always starts $a$ before $b$. It remains to choose $q$ such that the expected completion time of $T'$ is the same as $T$. We choose $q$ such that the probability of entering $T_R(a \rightarrow b)$ or $T_R(-a)$ is exactly $p_b$. We conflate the name of a subtree (e.g. $T_R(a \rightarrow b)$) with the event that we enter the subtree. Thus, we want $\mathbb{P}(T_R(a \rightarrow b) \vee T_R(-a)) = p_b$. The former probability is exactly $p_a + \bar{p}_a q p_b$, where we define $\bar{p} = 1 - p$ for a probability $p$. This gives $q = \frac{p_b - p_a}{\bar{p}_a p_b}$. Thus, we have chosen $q$ such that $\mathbb{P}(T_R(a \rightarrow b) \vee T_R(-a)) = p_b$ and $\mathbb{P}(T_L(a \rightarrow b) \vee T_L(-a)) = \bar{p}_b$. One should imagine that these two events are our replacements for the original tree $T$ entering $T_R$ and $T_L$. 
    
    In both subtrees $T_L(a \rightarrow b)$ and $T_L(-a)$, we replace the original job $a$ from $T_L$ with $b$ and a zero job, respectively. Let $\tilde{X}_a$ denote the size of the the replacement job, which is supported on $\{0, s\}$. We compute the distribution of $\tilde{X}_a$:
    \[\mathbb{P}(\tilde{X}_a = s \mid T_L(a \rightarrow b) \vee T_L(-a)) = \frac{\mathbb{P}(T_L(a \rightarrow b))}{\mathbb{P}(T_L(a \rightarrow b) \vee T_L(-a))} p_b = \frac{\bar{p}_a \bar{q}}{\bar{p}_b} p_b = p_a.\]
    It follows, conditioned on $T_L(a \rightarrow b) \vee T_L(-a)$, our replacement job for $a$ has the same distribution as $a$. An analogous computation for the right subtree gives:
    \[\mathbb{P}(\tilde{X}_a = s \mid T_R(a \rightarrow b) \vee T_R(-a)) = \frac{\mathbb{P}(T_R(a \rightarrow b))}{\mathbb{P}(T_R(a \rightarrow b) \vee T_L(-a))} p_b = \frac{p_a}{p_b} p_b = p_a,\]
    so the distribution of our replacement job conditioned on $T_R(a \rightarrow b) \vee T_R(-a)$ has the same distribution as $a$ as well.
    
    To summarize, we have constructed a tree $T'$ that starts $a$ before $b$. $T'$ enters $T_L(a \rightarrow b)$ or $T_L(-a)$ with probability $\bar{p}_b$: exactly the same as the probability that $T$ enters $T_L$. Further, $T'$ enters $T_L(a \rightarrow b)$ or $T_L(-a)$ with the same initial loads as $T$ entering $T_L$, because both correspond to all previous jobs in the subtree having size $0$. Finally, upon entering $T_L(a \rightarrow b)$ or $T_L(-a)$, the job we replace $a$ with has the same distribution as $a$. The analogous properties hold for the right subtree as well. We conclude, for any job $j \in J \setminus \{a,b\}$, the expected completion time of $j$ in $T'$ is the same as in $T$ (subject to the same initial loads.)
    
    It remains to show that the expected completion time of $a$ and $b$ weakly decreases from $T$ to $T'$. We define $\ell' \sim T_L$ to be the load of the least-loaded machine upon reaching the node $a$ in subtree $T_L$ (we define $\ell' \sim T_R$ analogously.) This is well-defined, because $a$ is scheduled on every root-leaf path in $T_L$. Note that $\ell'$ does not depend on the job scheduled at node $a$. It follows, the expected completion time of $a$ and $b$ in $T'$ are:
    \[\mathbb{E}_{T'} \,C_a = \ell + s p_a\]
    \[\mathbb{E}_{T'} \,C_b = \mathbb{P}(T_L(a \rightarrow b)) (\mathbb{E}_{\ell' \sim T_L} \ell' + sp_b) + \mathbb{P}(T_L(-a)) \ell + \mathbb{P}(T_R(-a)) (\ell + s) + \mathbb{P}(T_R(a \rightarrow b)) (\mathbb{E}_{\ell' \sim T_R} \ell' + s p_b).\]
    Now we simplify the completion time of $b$. First, we consider the terms corresponding to the left subtree. We have $\mathbb{P}(T_L(a \rightarrow b)) = \bar{p}_b\frac{p_a}{p_b}$, $\mathbb{P}(T_L(a \rightarrow b)) + \mathbb{P}(T_L(-a)) = \bar{p}_b$, and $\ell' \geq \ell$ for $\ell' \sim T_L$. Combining these three observations:
    \begin{align*}
        \mathbb{P}(T_L(a \rightarrow b)) (\mathbb{E}_{\ell' \sim T_L} \ell' + sp_b) + \mathbb{P}(T_L(-a)) \ell &= \mathbb{P}(T_L(a \rightarrow b)) \mathbb{E}_{\ell' \sim T_L} \ell' + \bar{p}_b s p_a + \mathbb{P}(T_L(-a)) \ell\\
        &\leq \bar{p}_b (\mathbb{E}_{\ell' \sim T_L} \ell' + s p_a).
    \end{align*}
    Now we consider the right subtree. Analogously, we have $\mathbb{P}(T_R(a \rightarrow b)) = p_a$, $\mathbb{P}(T_R(a \rightarrow b)) + \mathbb{P}(T_R(-a)) = p_b$, and $\ell' \geq \ell$ for $\ell' \sim T_R$. We compute:
    \begin{align*}
        \mathbb{P}(T_R(-a)) (\ell + s) + \mathbb{P}(T_R(a \rightarrow b)) (\mathbb{E}_{\ell' \sim T_R} \ell' + s p_b) &= (p_b - p_a) (\ell + s) + p_a \mathbb{E}_{\ell' \sim T_R} \ell' + s p_b + p_a s p_b\\
        &\leq p_b (\mathbb{E}_{\ell' \sim T_R} \ell' + p_a s) + (p_b - p_a) s.
    \end{align*}
    Combining our expressions for the left- and right-subtrees gives our final bound on the completion time of $a$ and $b$:
    \[\mathbb{E}_{T'} \,C_a + \mathbb{E}_{T'} \,C_b \leq \ell + s p_b
      + \bar{p}_b (\mathbb{E}_{\ell' \sim T_L} \ell' + s p_a) + p_b
      (\mathbb{E}_{\ell' \sim T_R} \ell' + p_a s) = \mathbb{E}_T \,C_b
      + \mathbb{E}_T \,C_a. \qedhere\]
\end{proof}

\section{Justification for \Cref{as_rescale}}\label{sec_appendix_rescale}

\rescale*
\begin{proof}
    Let $\mathcal{A}$ be the algorithm assumed by the lemma. We will run $\mathcal{A}$ on a subinstance of jobs satisfying \Cref{as_rescale}. Suppose we have a collection $J$ of Bernoulli jobs of the form $X_j \sim s_j \cdot Ber (p_j)$ for arbitrary size parameters $s_j$.
    
    First, we round up all size parameters to the nearest power of $2$. This at most doubles $\opt$. Then, we rescale all $s_j$'s uniformly so that $\sum_j \mathbb{E} X_j = 1$. Note that now we have $\mathbb{E} \opt \geq \sum_j \mathbb{E} X_j = \Omega(1)$. Finally, we partition $J = S \cup M \cup L$ into small, medium, and large jobs, respectively such that $S$ consists of the jobs $j$ with $s_j < \frac{1}{n^2}$, $M$ the jobs $j$ with $\frac{1}{n^2} \leq s_j < n^8$, and $L$ the jobs $j$ with $s_j \geq n^8$. Thus, $M$ is a collection of Bernoulli jobs satisfying \Cref{as_rescale}.
    
    Our algorithm to schedule $J$ is the following:
    \medskip
    \hrule
    \smallskip
    \hrule
    \medskip
    \begin{OneLiners}
        \item[i.] List-schedule all large jobs $L$ in arbitrary order.
        \item[ii.] List-schedule all small jobs $S$ in arbitrary order.
        \item[iii.] Run $\mathcal{A}$ to schedule the medium jobs $M$.
    \end{OneLiners}
    \medskip
    \hrule
    \smallskip
    \hrule
    \medskip
    It is clear that this algorithm is efficient, deterministic, and outputs a list schedule as long as $\mathcal{A}$ does as well. It remains to bound the total completion time of this schedule, which we denote by $\alg$. We let $B$ be the event that some large job comes up heads (i.e. has realized size at least $n^8$.)
    
    On the event $\bar{B}$, every large job comes up tails, so they contribute $0$ to $\alg$. Then we list-schedule the small jobs with initial load $0$ on every machine. The total completion time of all jobs in $S$ can be crudely upper-bounded by the max load after $S$ times the number of jobs, which is at most $\frac{1}{n} \cdot n = O( \mathbb{E} \opt)$.
    
    After this, we schedule the medium jobs using $\mathcal{A}$. After scheduling $S$, all machines are free by time $\frac{1}{n}$. Let $\mathcal{A}$ be the total completion time of running $\mathcal{A}$ on jobs $M$ starting at time $0$. We need the following monotonicity property of list schedules, which is analogous to \Cref{lem_listprop}
    
    \begin{lem}\label{lem_listprop_c}
		Consider a set of deterministic jobs and a fixed list schedule of those jobs. Then increasing the initial load or decreasing the number of machines weakly increase the total completion time of the schedule.
	\end{lem}
	\begin{proof}
		Let $J$ be the set of jobs. Consider initial load vectors $\ell, \ell' \in \mathbb{R}^m$, where the $i$th entry of each vector denotes the initial load on machine $i$. Now suppose $\ell \leq \ell'$, entry-wise. It suffices to show that $C(J,\ell) \leq C(J,\ell')$, where $C(J,\ell)$ is the total completion time achieved by our list-schedule with initial load $\ell$. This suffices, because we can decrease the number of machines by making the initial loads of some machines arbitrarily large so that they will never be used.
		
		We prove $C(J,\ell) \leq C(J,\ell')$ by induction on the number of jobs, $\lvert J \rvert$.
		In the base case, $\lvert J \rvert = 0$, so the claim is trivial because $C(J,\ell) = 0$ and $C(J,\ell') = 0$.
		For $\lvert J \rvert > 0$, let $j$ be the first job in the list, which is scheduled, without loss of generality, on the first machine for both initial loads $\ell$ and $\ell'$. Then:
		\[C(J,\ell) =  (\ell_1 + s_j) + C(J \setminus \{j\}, \ell + s_j e_1) \leq (\ell'_1 + s_j) + C(J \setminus \{j\}, \ell' + s_j e_1) = C(J,\ell'),\]
		where $e_1$ is the first standard basis vector, so we have $\ell + s_j e_1 \leq \ell' + s_j e_i1$ entry-wise. Then we assumed inductively that $C(J \setminus \{j\}, \ell + s_j e_1) \leq C(J \setminus \{j\}, \ell' + s_j e_1)$.
	\end{proof}
    
    By the above lemma, we can upper-bound the total completion time of $\mathcal{A}$ on jobs $M$ by starting once all machines are free after scheduling $S$, so at time $\frac{1}{n}$. This increases the completion time of each job by $\frac{1}{n}$. To summarize, on the event that every large job comes up tails, we have:
    \[\mathbb{E} \alg \cdot 1_{\bar{B}} \leq \frac{1}{n} \cdot n + \frac{1}{n} \cdot n + \mathbb{E} \mathcal{A} = O(\mathbb{E} \opt) + \alpha (\mathbb{E} \opt + O(1)) = O(\alpha) \cdot \mathbb{E} \opt,\]
    where we used the guarantee of $\mathcal{A}$ and $\mathbb{E} \opt = \Omega(1)$.
    
    It remains to consider the event where some large job comes up heads. In this case, we will not use the guarantee of $\mathcal{A}$. Instead, we will upper bound $\alg$ by the cost of an arbitrary list schedule. We define $S_1 = \max_{j \in J} X_j$ and $S_2$ to be the size of the second-largest job in $J$. On the event $B \cap \{S_2 \leq \frac{1}{n^2} S_1\}$, we note that no job is scheduled after the largest job with size $S_1$ on the same machine (using $m \geq 2$.) Noting all other jobs have size at most $\frac{1}{n^2} S_1$, we can upper bound $\alg$ by:
    \[\alg \leq S_1 + n \cdot \frac{1}{n} S_1 \leq 2 S_1 \leq 2 \opt,\]
    so we have $\mathbb{E} \alg \cdot 1_{B, S_2 \leq \frac{1}{n^2} S_1} = O(\mathbb{E} \opt)$.
    
    Finally, we bound $\mathbb{E} \alg \cdot 1_{B, S_2 > \frac{1}{n^2} S_1}$. We partition $B = \cup_{k = 0}^\infty B_k$, where $B_k = \{\max_{j \in J} X_j \in [2^k n^8, \, 2^{k+1} n^8)\}$. On the event $B_k \cap \{S_2 > \frac{1}{n^2} S_1\}$, there are at least two jobs of size at least $2^k n^6$. Recall that $\sum_{j \in J} \mathbb{E} X_j = 1$, so in particular $\mathbb{E} X_j \leq 1$ for all $j \in J$. Thus by Markov's inequality, $\mathbb{P}(X_j \geq 2^k n^6) \leq 2^{-k} n^{-6}$ for all $j \in J$. By union-bounding over all pairs of jobs in $J$:
    \[\mathbb{P}(B_k, \, S_2 > \frac{1}{n^2} S_1) \leq \mathbb{P}(\text{$\exists$ two jobs in $J$ with size at least $2^k n^6$}) \leq O(n^2) (2^{-k} n^{-6})^2.\]
    Further, on the event $B_k \cap \{1_{B, S_2 > \frac{1}{n^2} S_1}\}$, every job has size at most $2^{k+1} n^8$, so we have $\alg \leq n \cdot n 2^{k+1}n^8 = 2^{k+1} n^{10}$. Thus, for each $k$, we have:
    \begin{align*}
        \mathbb{E} \alg \cdot 1_{B_k, S_2 > \frac{1}{n^2} S_1} &\leq 2^{k+1} n^{10} \cdot \mathbb{P}(B_k, \, S_2 > \frac{1}{n^2} S_1)\\
        &= 2^{k+1} n^{10} \cdot O(n^2) (2^{-k} n^{-6})^2 = O(2^{-k}).
    \end{align*}
    To complete the proof, we partition $B = \cup_{k = 0}^\infty B_k$ to bound $\mathbb{E} \alg \cdot 1_{B, S_2 > \frac{1}{n^2} S_1}$:
    \[\mathbb{E} \alg \cdot 1_{B, S_2 > \frac{1}{n^2} S_1} = \sum_{k =
        0}^\infty  \mathbb{E} \alg \cdot 1_{B_k, S_2 > \frac{1}{n^2}
        S_1} = O(\sum_{k = 0}^\infty 2^{-k}) = O(\mathbb{E} \opt). \qedhere\]
  \end{proof}


\section{Concentration arguments}\label{sec_appendix_concentration}

    We need the following standard Chernoff bound.
    
    \begin{prop}[Chernoff bound]\label{prop_chernoff}
        Let $X = X_1 + \dots X_n$ be a sum of independent, $\{0,1\}$-valued random variables and $\mu = \mathbb{E} X$. Then we have:
        \begin{itemize}
            \item $\mathbb{P}(X \leq (1 - \delta) \mu) \leq \exp{\big(-\frac{\delta^2 \mu}{2}\big)}$ for all $0 \leq \delta \leq 1$.
            \item $\mathbb{P}(X \geq (1 + \delta) \mu) \leq \exp{\big(-\frac{\delta^2 \mu}{2 + \delta}\big)}$ for all $0 \leq \delta$.
        \end{itemize}
    \end{prop}

\bigjobsmallexpectation*
\begin{proof}
    Fix $c \geq 0$ which we will choose sufficiently large later. Then assume for contradiction that there exists a batch $k$ and threshold $\tau > 2 \mathbb{E} F^*(n-n/2^k)$ such that $\mathbb{E} \lvert J_k(> \tau) \rvert > m + c \sqrt{m}$.
    
    To reach a contradiction, it suffices to show that $\mathbb{P}(\lvert J_k(> \tau) \rvert \leq m) < \frac{1}{2}$. This is because on the complement event $\lvert J_k(> \tau) \rvert > m$ (which we assume happens with probability strictly larger than $\frac{1}{2}$), we also have $\lvert J^*_k(> \tau) \rvert > m$ by \Cref{lem_batch_prefix}. This implies $F^*(n - n/2^k) > \tau \geq 2 \mathbb{E} F^*(n-n/2^k)$. This would contradict the definition of $\mathbb{E} F^*(n-n/2^k)$.
    
    For convenience, let $\mu = \mathbb{E} \lvert J_k (> \tau) \rvert$. By Chernoff, we have:
    \[\mathbb{P}(\lvert J_k(> \tau) \rvert \leq m) = \mathbb{P}(\lvert J_k(> \tau) \rvert \leq \mu(1 - \frac{\mu - m}{\mu})) \leq \exp\big(-\frac{(\mu - m)^2}{2\mu}\big).\]
    There are two cases to consider. Recall that by assumption, we have $\mu > m + c \sqrt{m}$. If $\mu \geq 2m$, then $\mathbb{P}(\lvert J_k(> \tau) \rvert \leq m) \leq \exp{(- \frac{\mu}{8})} \leq \exp{(- \frac{m}{4})} < \frac{1}{2}$ for $m = \Omega(1)$ sufficiently large. Otherwise, $m + c \sqrt{m} < \mu < 2m$. Then $\mathbb{P}(\lvert J_k(> \tau) \rvert \leq m) \leq \exp{(- \frac{c^2 m}{2m})} < \frac{1}{2}$ for $c = O(1)$ sufficiently large.
\end{proof}

\bigjobchernoff*
\begin{proof}
    Note that there are $O(\log n)$ choices for $k$ and $L = O(\log n)$ relevant choices for $\tau$. Thus, by a standard union bound argument it suffices to show that for fixed $k$ and $\tau > 2 \mathbb{E} F^*(n-n/2^k)$, we have:
    \[\mathbb{P}(\lvert \lvert J_k(> \tau) \rvert - \mathbb{E} \lvert J_k(> \tau) \rvert \rvert > \Delta) = \frac{1}{\poly(n)}.\]
    
    Now we may assume $m$ is large enough so that $\mathbb{E} \lvert J_k(> \tau) \rvert \leq m + c \sqrt{m} \leq (c+1)m$ for sufficiently large constant $c \geq 0$ (guaranteed by \Cref{lem_big_job_small_expectation}.) Then we can bound the deviation of $\lvert J_k(> \tau) \rvert$ again with a Chernoff bound. Let $\mu = \mathbb{E} \lvert J_k(> \tau) \rvert$. We take $\Delta = O(\sqrt{\mu} \log n) = O(\sqrt{m} \log n)$.
    
    There are two cases to consider. If $\mu < \Delta$, then the lower tail is trivial:
    \[\mathbb{P}(\lvert J_k(> \tau) \rvert \leq \mu - \Delta) \leq \mathbb{P}(\lvert J_k(> \tau) \rvert < 0) = 0.\]
    For the upper tail we use Chernoff:
    \[\mathbb{P}(\lvert J_k(> \tau) \rvert \geq \mu + \Delta) = \mathbb{P}(\lvert J_k(> \tau) \rvert \geq (1 + \frac{\Delta}{\mu})\mu) \leq \exp{(- \frac{\Delta^2}{2\mu + \Delta})} \leq \exp{(- \frac{\Delta^2}{3\Delta})}  = \frac{1}{\poly(n)}.\]
    Otherwise, $\mu \geq \Delta$, so in particular $\frac{\Delta}{\mu} \leq 1$. Then we use Chernoff for both the lower- and upper tails:
    \[\mathbb{P}(\lvert J_k(> \tau) \rvert \leq \mu + \Delta) = \mathbb{P}(\lvert J_k(> \tau) \rvert \leq (1 + \frac{\Delta}{\mu})\mu) \leq \exp{(- \frac{\Delta^2}{2\mu})} = \frac{1}{\poly(n)}.\]
    \[\mathbb{P}(\lvert J_k(> \tau) \rvert \geq \mu + \Delta) = \mathbb{P}(\lvert J_k(> \tau) \rvert \geq (1 + \frac{\Delta}{\mu})\mu) \leq \exp{(- \frac{\Delta^2}{2\mu + \Delta})} \leq \exp{(- \frac{\Delta^2}{3\mu})}  = \frac{1}{\poly(n)}.\]
    
\end{proof}

\end{document}